\documentclass{article}
\usepackage[utf8]{inputenc}
\usepackage{amsmath}
\usepackage{amssymb}
\usepackage{graphicx}
\usepackage{caption}
\usepackage{float}
\usepackage{algorithmic}
\usepackage{algorithm}
\usepackage{xcolor}
\usepackage{hyperref}
\DeclareMathOperator{\R}{\mathbb{R}}

\def\argmin{\mathop{\rm argmin}}
\def \E{\mathbb{E}}

\def\Wc{{\cal W}}

\def \E{\mathbb{E}}

\def\1{{\bf 1}}

\def \N{\mathbb{N}}

\def\argmin{\mathop{\rm argmin}}
\def\argmin_#1{\underset{#1}{\mathrm{argmin\, }}}

\newtheorem{Theorem}{Theorem}[section]
\newtheorem{Remark}[Theorem]{Remark}

\usepackage[sorting=cms,authordate,backend=biber,strict]{biblatex-chicago}
\begin{document}
\title{Deep learning for efficient frontier calculation in finance.
\thanks{This work is supported by  FiME, Laboratoire de Finance des March\'es de l'Energie}
}
\author{
Xavier \sc{Warin}
\footnote{EDF R\&D \& FiME \sf \href{mailto:xavier.warin at edf.fr}{xavier.warin at edf.fr}} 
}

\maketitle

\begin{abstract}
We propose deep neural network algorithms to calculate the efficient frontier in the Mean-Variance and Mean-CVaR portfolio optimization problems.
Starting with the Mean-Variance portfolio optimization problem in the Black Scholes framework, we first compare the analytical solution to the computed one and show the efficiency of the methodology in high dimension. 
Adding  additional constraints, we compare different formulations and, getting the same frontier,  show  that the no short-selling and no borrowing problem can be solved with all formulations.
Then a new projected feedforward network is shown to be able to deal with some local and global constraints on the weights of the portfolio while outperforming classical penalization methods.
We extend our numerical results to assets following the Heston model and show that the results obtained in the Black Scholes method still hold.
At last we give numerical results for the more difficult Mean-CVaR optimization problem getting realistic results but only for algorithms with a global resolution of the problem.

\end{abstract}
\vspace{5mm}
\noindent {\bf Key words:} Deep neural networks,  finance, Mean-Variance, Mean-CVaR, efficient frontier.
\vspace{5mm}
\section{Introduction}
Portfolio selection in order to achieve a given expected return given an accepted risk has been a subject of research for more than 60 years. The first problem studied was the one period Mean-Variance problem  in \cite{mark52}, \cite{mark59}. An analytic solution has been proposed first in \cite{merton1972analytic} for a positive covariance matrix and when short selling is allowed. It is only in \cite{li2000optimal} that the multi-period case has been solved by  a reformulation of the problem as a linear quadratic (LQ) one. The solution for the continuous case when short selling is allowed in a complete Black-Scholes market has been proposed in  \cite{zhou2000continuous} still based on the LQ reformulation. The case without short selling has been solved two years after in \cite{li2002dynamic}. Notice that, in this case, a solution is provided  if borrowing is allowed so that the investment in the bond is not constrained.
Then an extension with  randomization in coefficients is proposed in \cite{lim2004quadratic}, the  risk on the correlations is studied in \cite{chiu2014mean}, \cite{ismpha19}.
Very recently  results are obtained in \cite{jaber2020markowitz} supposing that the volatility is rough \cite{gatheral2018volatility} and follows an affine and quadratic Volterra model: the Mean-Variance problem is solved in some case as the  explicit solution of  Riccati
  backward stochastic differential equations.\\
  All these theoretical results are interesting but of limited use by practitioner as  borrowing  is  generally not used and other operational  constraints are added:  investors  first tend to limit the rebalancing of the portfolio (only achieved at  discrete dates) to limit transaction cost by imposing constraints on the variation of the investment weights in the assets composing the portfolio. Second, they generally impose  strategic views on the weights that are only allowed to stay into given limits. In this case, numerical methods are necessary. Using conventional PDE methods  \cite{wang2010numerical} have solved many constrained problems in the case of a single risky asset following a Black-Scholes dynamic. The same methodology in the case of an asset with jumps has been used in 
  \cite{dang2014continuous}. This kind of approach can only be used at most with two or three assets and the resolution of a  realistic  portfolio selection problem is out of reach. In order to  tackle the multi-dimensional problem with constraints, \cite{cong2016multi} have proposed two algorithms based on the LQ formulation: the first being based on pure forward simulations is suboptimal with constraints, while the second using a backward recursion is based on regressions such that only rather low dimensional cases can be solved.\\
  The use of the variance to evaluate the risk has been questioned by both practitioners and researchers as it both penalizes gain and loss. Numerous downside risk measures  penalizing losses or low gains have been proposed in the literature. Among them, Lower Partial Moments (LPM) which have been proposed  more than forty years ago in \cite{fishburn1977mean} rely on two parameters : the gamma $\gamma$ parameter named the "Benchmark" parameter is set by the investor and  the second one $q$ represents the risk attitude of the investor. 
  This risk model embeds a lot of classical models. For example the case $q=0$ corresponds to the safety rule of \cite{roy1952safety}, the case $q=1$ correspond to the expected regret of \cite{dembo1999practice}, $q=2$ corresponds to the semi-deviation below the Benchmark parameter or the semi-variance if the Benchmark parameter is set to the expected wealth.\\
  Another classical risk measure related is the CVaR introduced in \cite{rockafellar2000optimization} corresponding the expected loss below the VaR measure. As shown in \cite{rockafellar2000optimization}, CVaR calculation can be parametrized as the minimum  over a parameter $\alpha$ of a function value linked to a LPM risk measure with parameter $q=1$. Using this formulation, \cite{gao2017dynamic} studied the Mean-CVaR continuous case giving semi-analytical solutions to the Mean-CVaR problem when the wealth is bounded. Indeed the boundedness of the wealth or  the control as studied in \cite{miller2017optimal} is necessary as the general case is not well posed for downside risk measures as shown in \cite{jin2005continuous}: using this kind of risk measure, and without constraints, the investors tend to gamble more and more if the market is in bad shape, and investments in the most risky assets tend toward infinity. Therefore, the Mean-CVaR problem has to be solved with constraints or in discrete-time and numerical methods are necessary to optimize portfolio. A classical approach consists in using the auxiliary formulation proposed  by \cite{rockafellar2000optimization}  and in using  a gradient descend method on $\alpha$ as proposed in \cite{miller2017optimal}. However, this procedure is very time consuming and only can be solved in low dimension. \\
  In order to optimize portfolio with general Mean-Risk measure, Neural Networks appears to be an interesting choice. Neural networks are known to be able to approximate  function  in high dimension.
  Very recently, neural networks have been used in risk management   first in  \cite{buehler2019deep}. Some cases with constraints on the hedging products and some downside risk  measures have been studied in \cite{fecamp2019risk}. In both cases, results reported are promising. Cases of Asset Liability Management have been reported very recently in \cite{krabichler2020deep}.\\
  In this article, we show that neural networks are able to calculate very realistic  efficient frontier in the Mean-Variance case and the Mean-CVar case.\\
  The main findings developed  are the following ones:
  \begin{itemize}
  \item {\bf Neural networks are able to solve  continuous Mean Variance problems accurately in low and high dimension for the control}: using different formulations, we first solve the continuous Mean-Variance problem without constraints in the Black-Scholes model with a direct formulation and the LQ formulation. We introduce for each formulation two methods to approximate the efficient frontier : the first one approximates the frontier point by point while the second permits to evaluate the global frontier in a single calculation.
  In all the cases, the frontier is correctly approximated in dimension 4 and 20 by comparison to the analytical solution.\\
  Then we used the  global and point by point  local method and adapt the network to solve the problem when no short selling and borrowing are allowed. Although we do not have any analytical solution, all frontiers calculated are very similar indicating that they are correctly evaluated.
  \item {\bf Neural networks  are  able to deal with  operational constraints}:
  in order to calculate the frontier when  global and local constraints are added to the resolution we propose different formulations based on different penalization methods. We show that the global constraints are hard to satisfy by penalization and we introduce a new projected feedforward network which is able to deal with the fact that the weights in the portfolio are such that all of them are positive, that their sum is equal to one and that each weights are between given bounds. 
  Numerical methods still indicate that the border is correctly calculated.
  \item  {\bf Neural networks are able to deal with the Mean Variance problem with a state in high dimension}:
  the Black-Scholes case is interesting but is special as the state of the problem only involve the global wealth: then as the number of assets increases, only the dimension of the control increases.
  In a  section we use the Heston model \cite{heston1993closed} and show that the frontiers are still correctly calculated with or without constraints. As the state of the problem depends on the wealth and the variance of the assets, we have shown that we are able to solve a high dimension  problem for both the state and the control. 
  \item {\bf Neural networks  can solve  Mean-CVaR problems accurately}: we solve some more difficult Mean-CVaR problems and show that on some cases, solutions are sometimes trapped in local  minima that can be far from the optimum. In this case,  multiple calculations can be achieved to get back the correct frontier using a   global formulation while the point by point formulation always gives oscillations.
  \end{itemize}
  We also show that, depending on the problem, one formulation may be preferred:
  \begin{itemize}
      \item For the Mean-Variance problem, a point by point approximation of the frontier is the best choice for the Black-Scholes model, while the global formulation is the best choice for the Heston model.
      \item Whereas for the Mean-CVaR, the global approaches   are the only  tested approaches to always obtain good results in the high dimension case (or what seems to be good results because no reference is available).
  \end{itemize}
  The article is organized as follows: In the first section, we briefly recall what a neural network is,  the gradient descent methodology used, the different choices of some hyper parameters, and how the convergence  of the optimization is checked. In  the second section, we solve the continuous Mean-Variance problem. Some constraints are added in the third section.
  Section four and five focus respectively on the use of the Heston model in the Mean-Variance setting and the Mean-CVaR setting with a Black-Scholes model.\\
  In the whole sequel, we note $(\Omega, \mathcal{F}, \mathbb{P},\mathcal{F}_t) $  a filtered probability space. For each set $\mathcal{A}$ in $\R^d$, we note $\mathcal{L}_{\mathcal{F}_t}^2(0,T,\mathcal{A})$  the set of the $\mathcal{F}_t$ adapted square integrable  processes  with values in $\mathcal{A}$.
  At last we suppose that the risk free rate is $0$ which is equivalent to discount all asset values with a rate  $r$, so that we  consider all risky assets with an adapted trend equal to the surplus of trend with respect to the non-risky asset.
  \paragraph{Notations:} In the sequel, for an element $x$ of $\R^d$,  $x^{+}= \max(x, 0)$ applied component by component,  for $(x,y)\in \R^2$, $x \vee y$ stands for $\max(x,y)$ and $x \wedge y$ stands for $\min(x,y)$.
 
\section{Neural networks as function approximators}
\label{secNN}
Deep neural  networks are designed to approximate  large class of functions. They   rely on the composition of simple functions, and appear to provide an efficient way to handle high-dimensional approximation problems, by finding the ``optimal" parameters by stochastic gradient descent methods. 
We here use a basic type of network dubbed feedforward networks. We fix the input dimension $d_0$ $=$ $d$ that will represent the dimension of the state variable $x$, the output dimension $d_1$ (here $d_1$ $=$ $1$  for approximating the real-valued solution to the PDE, or $d_1$ $=$ $d$ for approximating the vector-valued gradient function), the global number $L+1$ $\in$ $\N\setminus\{1,2\}$ of layers with $m_\ell$, $\ell$ $=$ $0,\ldots,L$, the number of neurons (units or nodes) on each layer: the first layer is the input layer with $m_0$ $=$ $d$, the last layer is the output layer with $m_L$ $=$ $d_1$, and the $L-1$ layers between are called hidden layers, where we choose for simplicity the same  dimension $m_\ell$ $=$ $m$, $\ell$ $=$ $1,\ldots,L-1$.
A  feedforward neural network is a  function from  $\R^{d}$ to $\R^{d_1}$ defined as the composition
\begin{align} \label{defNN}
x \in \R^d  & \longmapsto  \; A_L \circ  \varrho \circ A_{L - 1} \circ \ldots \circ \varrho \circ A_1(x) \; \in \; \R^{d_1}. 
\end{align}
Here $A_\ell$, $\ell$ $=$ $1,\ldots,L$ are affine transformations: $A_1$ maps from $\R^d$ to $\R^m$, $A_2,\ldots,A_{L-1}$ map from $\R^m$ to $\R^m$, and $A_L$ maps from $\R^m$ to 
$\R^{d_1}$, represented by 
\begin{align}
A_\ell (x) &= \; \Wc_\ell x + \beta_\ell,
\end{align}
for a matrix $\Wc_\ell$ called weight, and a vector $\beta_\ell$ called  bias term,  $\varrho$ $:$ $\R$ $\rightarrow$ $\R$ is a nonlinear function, called activation function, and applied 
component-wise on the outputs of $A_\ell$, i.e., $\varrho(x_1,\ldots,x_m)$ $=$ $(\varrho(x_1),\ldots,\varrho(x_m))$. Standard examples of activation functions are 
the sigmoid, the ReLU, the ELU, $\tanh$. 
Generally, for stochastic control problems, the number of layers is kept low (between 2 and 4) in order to avoid the problem of vanishing gradient and the number of neurons depends on $d$  but is kept generally between $10$ and $100$ (\cite{warin2021reservoir},\cite{chan2019machine}, \cite{fecamp2019risk}, \cite{han2018solving}).

All these matrices $\Wc_\ell$ and vectors $\beta_\ell$, $\ell$ $=$ 
$1,\ldots,L$,  are the parameters of the neural network, and can be identified with  an element $\theta$ $\in$ $\R^{\kappa_{L,m}}$, where $\kappa_{L,m}$ $=$ 
$\sum_{\ell=0}^{L-1} m_\ell (1+m_{\ell+1})$ $=$ $d(1+m)+m(1+m)(L-2)+m(1+d_1)$ is the number of parameters, where we fix $d_0$, $d_1$, $L$, and $m$.  
The fundamental result of Hornick et al. \cite{horetal89} justifies the use of neural networks as function approximators by proving that the set of all feedforward
networks letting $m$ vary is dense in  $L^2(\nu)$ for any finite measure $\nu$ on $\R^d$, whenever $\varrho$ is continuous and non-constant. \\
In the whole article, we use  three hidden layers and a number of neurons equal to $10+d$ such that the dimensional space for the parameters of the neural network is $\hat \kappa = \kappa_{4,10+d}$. Gradient descent is implemented in Tensorflow \cite{2015tensorflow} using ADAM optimized \cite{kingma2014adam}.
The  learning rate used by Adam is taken linearly decreasing with gradient iterations from a given initial value to a final one. 
The activation function used is the tanh function which has been widely used for the resolution of non linear  PDEs \cite{hure2020deep} \cite{pham2021neural} \cite{germain2020deep}. Compared to ReLU activation functions it gives slightly better results and, as it is bounded, gradient descent algorithms diverge less easily for high learning rates.
In the sequel the batch size is generally chosen between $100$ and $300$. Other different parameters such as the learning rates and the number of gradient iterations are chosen such that the solution obtained does not vary a  lot with iterations : to check that the number of  iterations is sufficient, every $100$ gradient iterations, an accurate estimation of the objective function is achieved with a high number of samples permitting to check that the objective function value is stabilized.

\section{Mean-Variance efficient frontier in a continuous setting}  
In this section we suppose that the asset prices follow a Black-Scholes model:
\begin{align}
    \frac{dS_t}{S_t} =  \mu dt +\sigma dW_t 
    \label{eq:BS}
\end{align}
with $S_t$ with values $\R^d$ with components $S_{t,j}, \quad j=1, \dots, d$, $\mu$ with values in  $\R^d$, $\sigma \in \{ \mbox{diag}(v), v \in \R^d_{> 0} \}$ the set of  diagonal matrices with strictly positive values, $W_t= (\hat W^i_t)_{i=1,d}$ where the $\hat W^i$ are $\mathcal{F}_t$  adapted  Brownian motions correlated with a correlation matrix $\rho$.\\
In the continuous setting, we note $\xi = (\xi_t)_{t>0}$  the investment strategy with values in  $\R^d$ until maturity $T$, with a component $i$  corresponding to the fraction of wealth invested in asset $i$.  We suppose $\xi_t$ is in  $\mathcal{L}_{\mathcal{F}_t}^2(0,T,\R^d)$. 
The portfolio value at date $T$ verifies:
\begin{align}
    X_T^{\xi} = X_0 + \int_0^T \xi_t X_t^{\xi}. \frac{dS_t}{S_t} = X_0 + \int_0^T X_t^{\xi} \xi_t. (\mu dt + \sigma dW_t)
    \label{eq:conPorfolio}
\end{align}
The Mean-Variance problem consist in finding strategies $\xi$ adapted to the available information that minimize:
\begin{align}
    (J_1(\xi), J_2(\xi)) := ( -\E[ X_T^{\xi}], \E[(X_T^{\xi}-\E[X_T^{\xi}])^2]).
\end{align}
An admissible strategy $\xi^{*}$  is said to be efficient if there is no other strategy $\psi$ such that
\begin{align*}
    J_1(\psi) \le J_1(\xi^{*}), \quad  J_2(\psi) \le J_2(\xi^{*})
\end{align*}
and at least one of the two previous inequalities is strict.\\
Then  $( J_1(\xi^{*}), J_2(\xi^{*})) $  is an efficient point and the set of all efficient points defines the efficient frontier. By convexity (see \cite{zhou2000continuous}), this Pareto frontier can be calculated by minimizing the function defined as the weighted average of the two criteria:
\begin{align}
    J_1(\xi) + \beta J_2(\xi)
    \label{eq:minimPro}
\end{align}
where the parameter  $\beta >0$  defines a point of the efficient frontier.\\
In the continuous setting, the solution of the Mean-Variance problem is known explicitly \cite{zhou2000continuous},\cite{ismpha19} and the optimal investment strategy 
 $ \alpha_t = \xi^{*}_t X^{\xi^{*}}_t$  where  $\xi^{*}$ minimizes \eqref{eq:minimPro} and  $
X^{\xi^{*}}_t$ is the optimal portfolio associated  is given by:
\begin{align}
    \alpha(X_t^{\xi^{*}}) =  - ( \sigma \rho \sigma)^{-1} \mu  \Big[ X^{\xi^{*}}_t -  X_0  - \frac{e^{RT}}{2\beta} \Big], \quad 0 \leq t \leq T
    \label{eq:optComCont}
\end{align}
where $R=  \mu. (( \sigma \rho \sigma)^{-1} \mu ) $.\\
The problem \eqref{eq:minimPro} does not admit any dynamic programming principal due the average term in the variance definition so that no PDE or regression method can be used directly to solve it. 
The derivation of the analytic solution is based on a LQ auxiliary equivalent formulation as shown in \cite{zhou2000continuous}. Indeed the solution of 
the problem \eqref{eq:minimPro} is solution of 
\begin{align}
 \xi^{*} = \argmin_{\xi \in \mathcal{L}_{\mathcal{F}_t}^2(0,T,\R^d) }   \E[(X_T^{\xi} -\gamma)^2]
 \label{eq:auxi}
\end{align}
where
\begin{equation}
 \gamma = \frac{1}{2\beta} + \E[X_T^{\xi^{*}}]  .
 \label{eq:gammaAux}
\end{equation}
It is then possible to estimate the efficient frontier by the resolution of the problem \eqref{eq:auxi} by letting $\gamma$ vary. This kind of formulation is generally used by conventional methods such as regressions \cite{cong2016multi} and PDEs \cite{wang2010numerical}, \cite{dang2014continuous} as the dynamic programming principle can be used.
 
 \subsection{Neural network approximations}
 Equation \eqref{eq:conPorfolio} is first discretized on grid of dates $t_i, 0 \le i < N$ such that $t_0= 0$,  $0 < t_i < T$ for $0< i < N$ and we note $t_N=T$. We note $\phi^i$ a position (as a fraction of the wealth invested in each asset) at date $t_i$, and $\phi = (\phi^i)_{i=0, \dots,N-1}$. The portfolio value is then given by
\begin{align}
    X_T^{\phi} = X_0 + \sum_{i=0}^{N-1} \phi^i X_{t_i}^{\phi}. \frac{S_{t_{i+1}}- S_{t_i}}{S_{t_i} }
    \label{eq:conPorfolioDis2}
\end{align}
We first present the methodology used to solve \eqref{eq:minimPro} for a given $\beta$ (or \eqref{eq:auxi}  for a given $\gamma$).
The state of the system only depends on $t$ and  the wealth $x$ as shown by equation \eqref{eq:conPorfolio}, then  we classically introduce a single network with parameters $\theta \in \R^{\hat \kappa}$ taking $t$ and the wealth $x$ as input (so in dimension 2) and with output $\hat \phi^{\theta}(t,x) $ in dimension $d$ where $\hat \phi_j^{\theta}(t_i,.)$ is an approximation of $\phi^i_j$. 
\begin{Remark}
In the general case, where the asset is not modeled by a geometric brownian motion,  the value of the asset has to be included in the state.
\end{Remark}
No activation function is used on the final output such that the network gives an output potentially covering  $\R^d$.
Then we solve
\begin{align}
   \theta^{*} = \argmin_{\theta \in \R^{\hat \kappa }} -\E[X_T^{\hat \phi^{\theta}}] + \beta \E[ (X_T^{\hat \phi^{\theta}}-\E[X_T^{\hat \phi^{\theta}}])^2]
   \label{eq:directNN}
\end{align}
for problem \eqref{eq:minimPro} or 
\begin{align}
   \theta^{*} =\argmin_{\theta \in \R^{\hat \kappa} }  \E[ (X_T^{\hat \phi^{\theta}}-\gamma)^2]
   \label{eq:ZhouForm}
\end{align}
for problem \eqref{eq:auxi}.\\
Instead of evaluating the frontier by solving \eqref{eq:directNN} for each $\beta$ chosen (or \eqref{eq:ZhouForm} for each $\gamma$), a second resolution method aims at evaluating the global frontier in one calculation.
In this case, taking for example problem \eqref{eq:minimPro}, we introduce a network with input  $(t,x, \beta)$ in dimension $3$ and still with an output in dimension $d$. The associated strategy is noted
$\hat \phi^{\theta; \beta}$ to emphasize the dependence on $\beta$ and we solve :
\begin{align}
   \theta^{*} = \argmin_{\theta \in \hat \kappa}    \E[-\E[X_T^{\hat \phi^{\theta; \hat \beta}}| \hat \beta] + \hat \beta \E[ (X_T^{\hat \phi^{\theta; \hat \beta}}-\E[X_T^{\hat \phi^{\theta; \hat \beta} }|\hat \beta])^2 |\hat \beta]]
   \label{eq:solvMeanVarBeta}
\end{align}
where  $E[ | \hat \beta]$  stands for the conditional expectation with respect to $\hat \beta$  which is a random variable with density that can be taken
\begin{itemize}
    \item either with discrete values  so $p(x) = \frac{1}{K}\sum_{j}^K  \delta_{\beta_j}(x)$ where  $(\beta_i)_{i=1,K}$ is a set of values where we want to approximate the frontier (generally $\beta_1 =0$) and in this case it is
    equivalent to minimize 
    \begin{align}
   \theta^{*} = \argmin_{\theta}  \sum_{k=1}^K  -\E[X_T^{\hat \phi^{\theta;  \beta_k}}] + \beta_k \E[ (X_T^{\hat \phi^{\theta;  \beta_k}}-\E[X_T^{\hat \phi^{\theta;  \beta_k}}])^2]
   \label{eq:solvMeanVarGam}
\end{align}
\item or $p(x)$ is for example an uniform law on $ [\underline \beta, \bar \beta]$  representing where we want to approximate the frontier.
\end{itemize}

All results in the following section are obtained using some batch of size $300$, the linear rate is taken linearly decreasing from $ 25 \times 10^{-4}$  to $25 \times 10^{-5}$ with gradient iterations. The number of iterations is set to $15000$. After training using $40$ points  ($K=40$ in the global estimation in equation \eqref{eq:solvMeanVarGam}, or $40$ points to approximate the frontier point by point), each point of the frontier is plotted calculating mean and variance using $1e5$ simulations.
\begin{Remark}
We use a single network for the resolution. It is also possible to use a neural network at each date as used in \cite{han2018solving} for the resolution of non linear PDEs. It turns out that the use of a single network as proposed in \cite{chan2019machine} for the same method highly improves the results in all  the tested configurations in this article. 
Note that there are real life problems in finance such as gas storage valuation  where the use of a single network is not sufficient \cite{warin2021reservoir}.
\end{Remark}
\subsection{Results in dimension 4}
We take $\mu = (0.01  ,0.0225, 0.035,  0.0475)^T,$ the diagonal of  $\sigma$ is given by $$(0.05, 0.1 , 0.15, 0.2 )$$ and we take
\begin{align*}
    \rho = \left (
    \begin{array}{cccc}
     1.   &       0.26 & -0.43 &  0.233 \\
  0.26  &  1. &          0.003 &   0.06 \\
 -0.43  & 0.003 &  1.       &  -0.33 \\
  0.233 &  0.06 & -0.33&  1.    \\
    \end{array}
    \right).
\end{align*}
We suppose that  $T=1$ year  and that the rebalancing is achieved twice a week  ($N=104$) to approach a continuous rebalancing.
We plot the frontier obtained  with $\beta$  values between $0.05$ to $2.7$.
The analytic solution is obtained by applying the continuous optimal control \eqref{eq:optComCont}  at each  rebalancing date.
In the sequel, in the figures, "Point by point" stands for an approximation using \eqref{eq:directNN}, "point by point auxiliary"
for an approximation using \eqref{eq:ZhouForm}, "global" stands for an approximation using \eqref{eq:solvMeanVarGam}, "global random" stands
for an approximation using \eqref{eq:solvMeanVarBeta} and an uniform law for $\hat \beta$. At last "global auxiliary" and "global auxiliary random"  stand  respectively for a global resolution with deterministic and random $\gamma$ values for the auxiliary problem \eqref{eq:auxi}.
 \begin{figure}[H]
    \centering
    \begin{minipage}[c]{.32\linewidth}
    \includegraphics[width=0.99\linewidth]{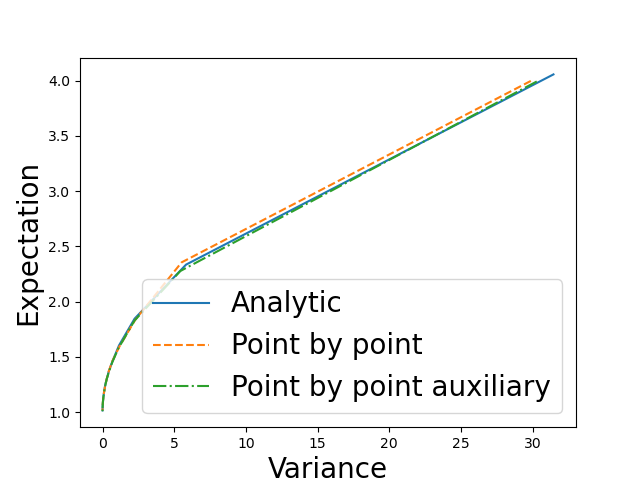}
    \caption*{Point by point}
    \end{minipage}
    \begin{minipage}[c]{.32\linewidth}  
    \includegraphics[width=0.99\linewidth]{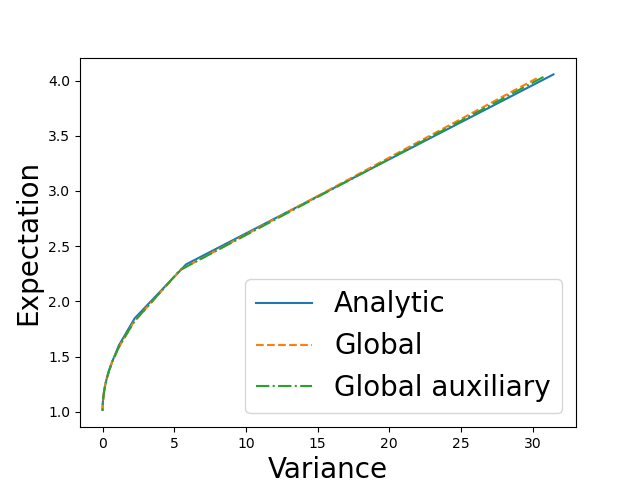}
    \caption*{Global}
    \end{minipage}
    \begin{minipage}[c]{.32\linewidth}  
    \includegraphics[width=0.99\linewidth]{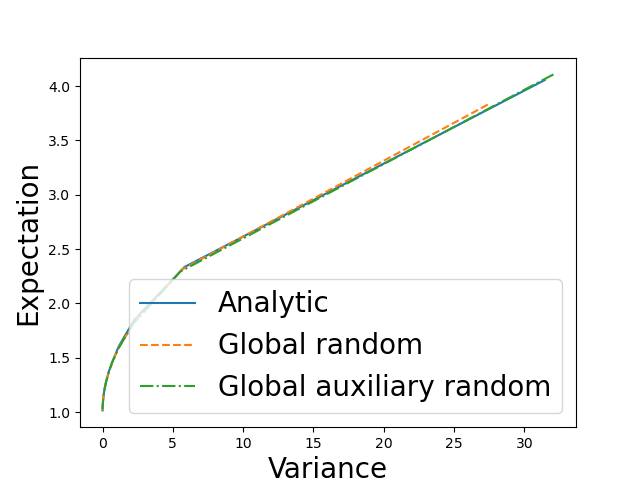}
    \caption*{Global random}
    \end{minipage}
    \caption{Efficient frontier in dimension 4.}
    \label{fig:BASDim4}
\end{figure}
Results on figure \ref{fig:BASDim4} show that the whole frontiers obtained are on the analytic frontier but global random estimation on the direct problem \eqref{eq:minimPro} tend to fail to reproduce the whole curve. Portfolio with very high returns are not found (the part of the curve with very high returns is missing). Nevertheless results are very good.
\begin{Remark}
For the smallest $\beta=0.05$ of the curve, the global random result obtained with the direct formulation  is rather unstable and  obtained doubling the learning rates. An increase of the number of  gradient iterations does not improve the results.
\end{Remark}

In table \ref{tab:someBetaDim4Direct} and \ref{tab:someBetaDim4Zhou}, we give the results obtained for 3 points of the curve corresponding to a low, a medium and a high variance result. Results are all very good except with the global random method associated with the direct approach   \eqref{eq:directNN}  for the high variance case, or associated to the auxiliary approach \eqref{eq:ZhouForm} for the low variance case.

\begin{table}[H]
    \centering
    \begin{tabular}{|c|c|c|c|c|c|c|c|c|} \hline
      $\beta$ &   \multicolumn{2}{c|}{Analytical} &  \multicolumn{2}{c|}{Point by point} &  \multicolumn{2}{c|}{Global} &  \multicolumn{2}{c|}{Global random} \\ \hline
 &  Mean & Variance  &  Mean & Variance  & Mean & Variance  & Mean & Variance   \\ \hline
0.05 & 4.056 & 31.445 & 3.997 & 29.868 & 4.021 & 30.281 & 3.83 & 27.4  \\ \hline
0.2 & 1.779 & 1.949 & 1.746 & 1.845 & 1.76 & 1.914 & 1.765 & 1.915  \\ \hline
2.0 & 1.077 & 0.019 & 1.08 & 0.021 & 1.078 & 0.02 & 1.075 & 0.018  \\ \hline 
\end{tabular}

    \caption{Resolution of \eqref{eq:directNN} for some $\beta$ values in a continuous setting in dimension 4.}
    \label{tab:someBetaDim4Direct}
\end{table}

\begin{table}[H]
    \centering
    \begin{tabular}{|c|c|c|c|c|c|c|c|c|} \hline
      $\gamma$ &   \multicolumn{2}{c|}{Analytical} &  \multicolumn{2}{c|}{Point by point} &  \multicolumn{2}{c|}{Global} &  \multicolumn{2}{c|}{Global random} \\ \hline
 &  Mean & Variance  &  Mean & Variance  & Mean & Variance  & Mean & Variance   \\ \hline
 14.097 & 4.107 & 30.324 & 3.992 & 29.941 & 4.033 & 30.748 & 4.108 & 32.068  \\ \hline
4.274 & 1.778 & 1.951 & 1.758 & 1.889 & 1.749 & 1.863 & 1.822 & 2.882  \\ \hline
1.327 & 1.077 & 0.019 & 1.075 & 0.019 & 1.077 & 0.021 & 1.099 & 0.906  \\ \hline
   \end{tabular}
    \caption{Resolution of \eqref{eq:ZhouForm} for some $\gamma$ values in a continuous setting in dimension 4.}
    \label{tab:someBetaDim4Zhou}
\end{table}
\subsection{Results in dimension 20}
\begin{figure}[H]
    \centering
    \begin{minipage}[c]{.32\linewidth}
    \includegraphics[width=0.99\linewidth]{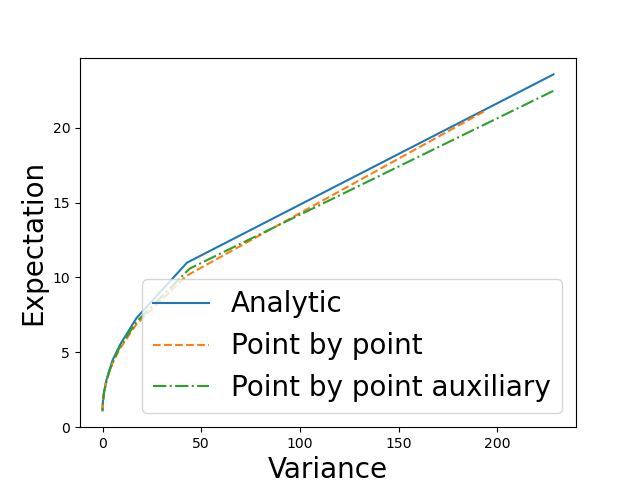}
    \caption*{Point by point}
    \end{minipage}
    \begin{minipage}[c]{.32\linewidth}  
    \includegraphics[width=0.99\linewidth]{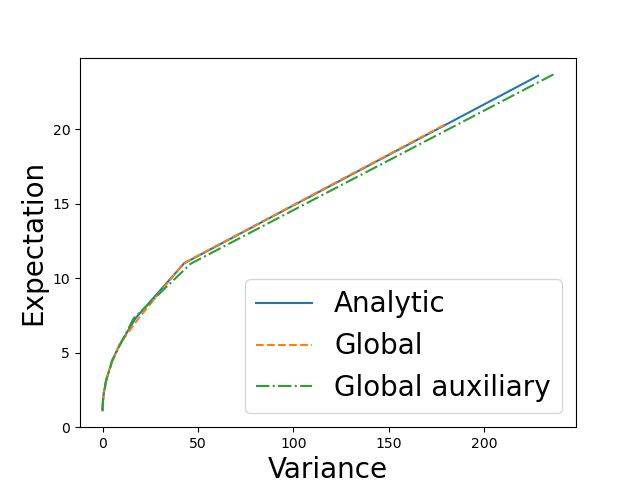}
    \caption*{Global}
    \end{minipage}
    \begin{minipage}[c]{.32\linewidth}  
    \includegraphics[width=0.99\linewidth]{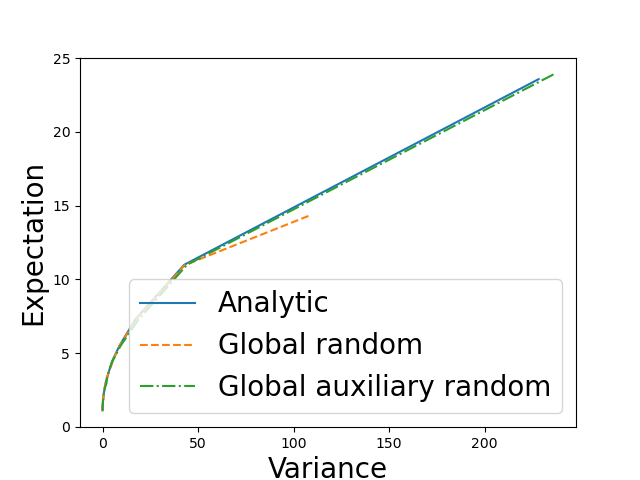}
    \caption*{Global random}
    \end{minipage}
    \caption{Efficient frontier in dimension 20.}
    \label{fig:BASDim20}
\end{figure}

We take $\mu_i=  0.01 + \frac{i-1}{400}$ and the diagonal matrix $\sigma$ such as  $\sigma_{i,i}= 0.05+ \frac{i-1}{100}$ for $i=1,\dots,20$.
The correlation is picked up randomly by a random generator but avoiding too high correlation leading to huge positions in the portfolio.
Results on figure \ref{fig:BASDim20} show that the point by point approximation especially  for the direct approach is less effective for very small $\beta$. The global approach can slightly outperform the analytic solution as the continuous formula is applied on a discrete-time  problem. 
\begin{Remark}
For the global random direct approach, the point (with high variance) on the right picture in figure \ref{fig:BASDim20} not aligned with  the points of the other curves  is not improved  while changing the hyper parameters.
\end{Remark}
As in dimension 4, in table \ref{tab:someBetaDim20Direct} and \ref{tab:someBetaDim20Zhou}, we give the results obtained for 3 points of the curve corresponding to a low, a medium and a high variance result.
\begin{table}[H]
    \centering
    \begin{tabular}{|c|c|c|c|c|c|c|c|c|} \hline
      $\beta$ &   \multicolumn{2}{c|}{Analytical} &  \multicolumn{2}{c|}{Point by point} &  \multicolumn{2}{c|}{Global} &  \multicolumn{2}{c|}{Global random} \\ \hline
 &  Mean & Variance  &  Mean & Variance  & Mean & Variance  & Mean & Variance   \\ \hline
0.05 & 24.204 & 238.117 & 21.094 & 193.111 & 20.278 & 179.205 & 14.301 & 107.822  \\ \hline
0.2 & 6.774 & 14.265 & 6.349 & 13.287 & 6.426 & 13.063 & 6.23 & 12.925  \\ \hline
2.0 & 1.583 & 0.145 & 1.541 & 0.13 & 1.562 & 0.142 & 1.564 & 0.142  \\ \hline
    \end{tabular}
    \caption{Resolution of \eqref{eq:directNN} for some $\beta$ values in a continuous setting in dimension 20. }
    \label{tab:someBetaDim20Direct}
\end{table}

\begin{table}[H]
    \centering
    \begin{tabular}{|c|c|c|c|c|c|c|c|c|} \hline
      $\gamma$ &   \multicolumn{2}{c|}{Analytical} &  \multicolumn{2}{c|}{Point by point} &  \multicolumn{2}{c|}{Global} &  \multicolumn{2}{c|}{Global random} \\ \hline
 &  Mean & Variance  &  Mean & Variance  & Mean & Variance  & Mean & Variance   \\ \hline
34.146 & 24.096 & 241.119 & 22.472 & 228.249 & 23.669 & 236.393 & 23.898 & 236.101  \\ \hline
9.286 & 6.802 & 13.695 & 6.556 & 14.524 & 6.731 & 14.641 & 6.971 & 18.877  \\ \hline
1.829 & 1.579 & 0.151 & 1.569 & 0.146 & 1.587 & 0.158 & 1.83 & 3.935  \\ \hline
   \end{tabular}
    \caption{Resolution of \eqref{eq:ZhouForm} for some $\gamma$ values in a continuous setting in dimension 20. }
    \label{tab:someBetaDim20Zhou}
\end{table}
Using results both in dimension 4 and 20, it seems that:
\begin{itemize}
    \item for high variance cases,  formulation \eqref{eq:ZhouForm} should be preferred,
    \item the global random method appears to have  difficulties to catch different extreme points of the curve depending on the formulation selected.
\end{itemize}
\section{Mean-Variance  on the discrete case adding constraints}
\label{sec:DisMeanVar}
In the section, we focus on the discrete case still trying to solve \eqref{eq:minimPro} or \eqref{eq:auxi}.
\begin{Remark}
We always suppose that the allocation weights belong to a convex set and then \eqref{eq:auxi} and \eqref{eq:minimPro}  remain equivalent.
\end{Remark}
In the whole section, we impose that there is no short selling, no borrowing and that all the whole  wealth is invested. Then in the two sections below, the $\phi^i$ in equation
\eqref{eq:conPorfolioDis2} satisfy:
\begin{align}
  0 \le \phi^i_j \le 1, & \quad \forall i=0, \dots, N-1, \quad j=1, \dots, d \nonumber \\
  \sum_{j=1}^d \phi^i_j  =1,  & \quad  \forall i=0, \dots, N-1
  \label{eq:constEasy}
\end{align}
In this case, we can rewrite equation \eqref{eq:conPorfolioDis2} by introducing  the yield vector $\mathbb{Y}_i = \frac{S_{t_{i+1}}- S_{t_i}}{S_{t_i} }$ so that:
\begin{align}
    X_T^{\phi} = X_0  \prod_{i=0}^{N-1} (1 +   \phi_i. \mathbb{Y}_i)
    \label{eq:conPorfolioDis}
\end{align}
In all the cases in this section we take $T=10$ years and we suppose that rebalancing is achieved once a month.
In dimension 4 and 20, trends and volatilities are the same as in the continuous case but correlations are again picked up randomly and for example in dimension 4:
\begin{align*}
    \rho = \left (
    \begin{array}{cccc}
    1.  &        0.805 & -0.894 &  0.59 \\
  0.805 &  1.  &        -0.571 &   0.473  \\
  -0.894 & -0.571&   1. &         -0.772 \\
 0.59  & 0.473 &   -0.772 &   1.  \\
    \end{array}
    \right).
\end{align*}
\begin{Remark}
No limits are imposed anymore on the correlation used as the weights are naturally bounded.
\end{Remark}
\subsection{Dynamic versus static optimization.}
The most interesting problem consists in finding the weights satisfying the equation \eqref{eq:constEasy} and resulting from strategies letting the weights vary with time and adapt to the state of the world.\\
Another  strategy widely used by the practitioners consists in  taking the weights constant leading to the  "constant mix" strategy. The value of the weights can be chosen by expert or optimized to minimize a given criterion.
Then  an  efficient frontier in the class of strategies  with constant weights can be calculated too.\\
In practice, engineers, trying to optimize  these constant weights,  draw randomly the weights and, for a set a weights drawn,  test the given strategy leading to a single point  (Variance, Expectation) using a very high  number of asset trajectories. When the number of assets is not too important, this brute force calculation permits to identify the sets of weights near the efficient frontier.\\
As the methodology is  easy to implement, this method is widely used. As the weights are constant, the rebalancing of  the portfolio  is generally limited and practitioners do not  have to justify  the high rebalancing of the portfolio that a dynamic strategy may propose.\\
In the sequel of the article, we will note by static optimization ("Static" on figures) the  optimal constant mix strategy such that  the weights are kept constant during the whole period. 
 When "Static" is not specified, a plot is carried out with a dynamic optimization.
\begin{Remark}
For a static optimization no neural network is used and the problem is reduced to an optimization in dimension $d$.
\end{Remark}
\subsection{No other constraints}
In this section, we suppose that only constraints \eqref{eq:constEasy} are imposed.
We use a similar network as in the previous section with parameter $\theta$ except that we use a sigmoid activation function  at the  output  giving a network $\zeta^\theta(t,X)$  for the point by point evaluation   with values in  $[0,1]^d$. Investment weights are then given by
\begin{align}
  \hat \phi^\theta(t,X) = \frac{\zeta^\theta(t,X)}{\displaystyle{\sum_{i=1}^d}\zeta^{\theta}_i(t,X)}  
  \label{eq:NNWeight1}
\end{align}
Then equation \eqref{eq:directNN} or \eqref{eq:ZhouForm} can be solved.\\
Similarly for a global formulation the network for the direct problem \eqref{eq:minimPro} take $(t,x,\beta)$ as input and the weights are defined as 
\begin{align}
  \hat \phi^{\theta;\beta}(t,X,\beta) = \frac{\zeta^\theta(t,X, \beta)}{\displaystyle{\sum_{i=1}^d} \zeta^{\theta}_i(t,X, \beta)}.  
  \label{eq:NNWeight1Glob}
\end{align}
Then it is possible to solve \eqref{eq:solvMeanVarBeta} or to minimize the objective function corresponding to the auxiliary problem.\\
In this section we take an initial learning rate equal to $\frac{1}{2}\frac{ 10^{-2}}{d}$ and linearly decreasing with iterations to $\frac{1}{2}\frac{10^{-3}}{d}$. The number of gradient iterations is set to $15000$ in dimension 4 and $30000$ in dimension 20, the batch size equal to $300$.
\begin{figure}[H]
    \centering
    \begin{minipage}[c]{.49\linewidth}
    \includegraphics[width=0.99\linewidth]{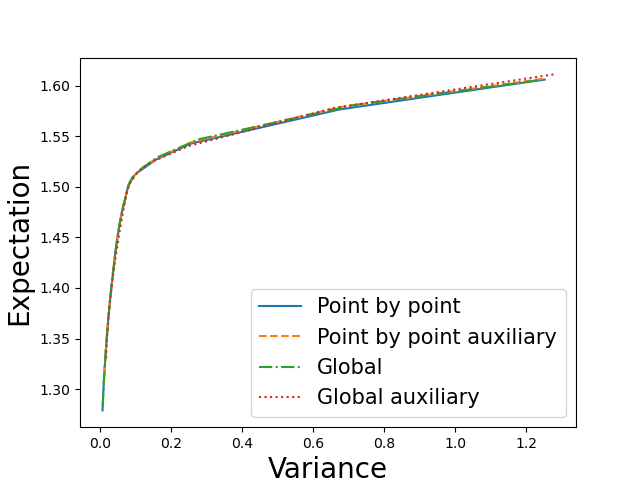}
    \caption*{Dimension 4}
    \end{minipage}
    \begin{minipage}[c]{.49\linewidth}  
    \includegraphics[width=0.99\linewidth]{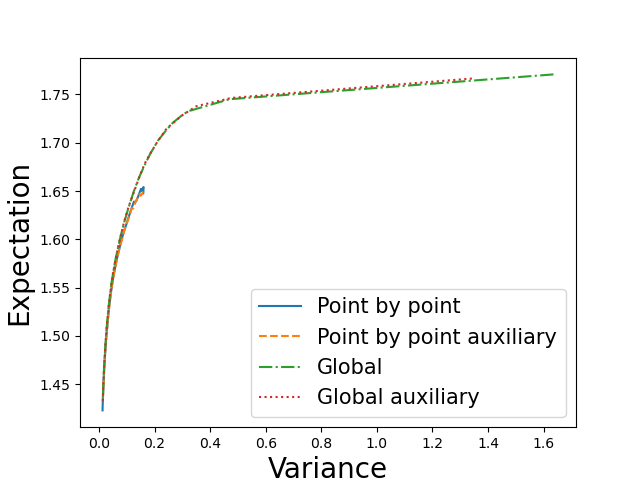}
    \caption*{Dimension 20}
    \end{minipage}     
    \caption{Efficient frontier with static optimization.}
    \label{fig:staticWeight01}
\end{figure}
On figure \ref{fig:staticWeight01}, we plot the efficient frontier calculated by static optimization for different methods.
In dimension 4, all methods seem to get back the whole frontier while in dimension 20  very high returns are hard to catch : the global  methods can get the part of the curve with very high returns  quite easily while point by point methods fail even when increasing the number of gradient iterations.
\begin{figure}[H]
    \centering
    \begin{minipage}[c]{.49\linewidth}
    \includegraphics[width=0.99\linewidth]{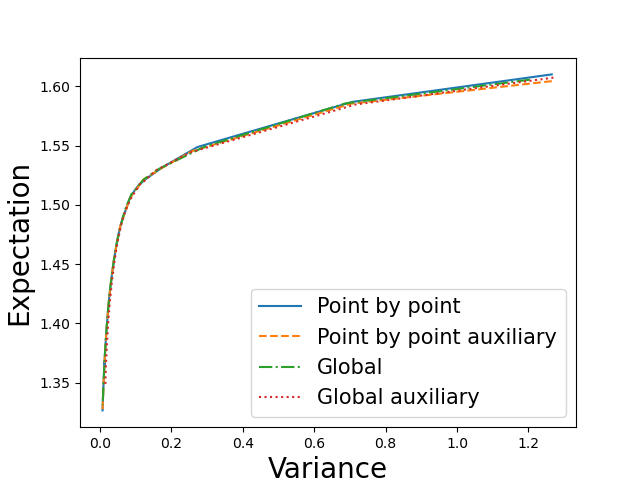}
    \caption*{Dimension 4}
    \end{minipage}
    \begin{minipage}[c]{.49\linewidth}  
    \includegraphics[width=0.99\linewidth]{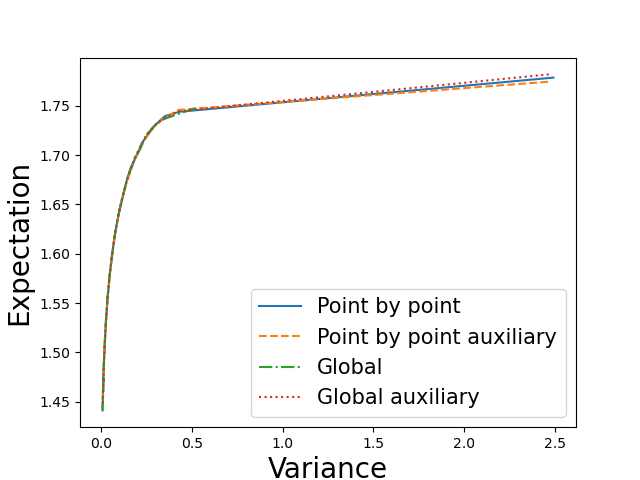}
    \caption*{Dimension 20}
    \end{minipage}     
    \caption{Efficient frontier with dynamic optimization  (point by point and global methods).}
    \label{fig:dynWeight01}
\end{figure}
As seen on figure \ref{fig:dynWeight01} dealing with the dynamic case, global with deterministic  $\beta$ (respectively $\gamma$) coefficients and point by point methods give very similar results.\\

\begin{figure}[H]
    \centering
    \begin{minipage}[c]{.49\linewidth}
    \includegraphics[width=0.99\linewidth]{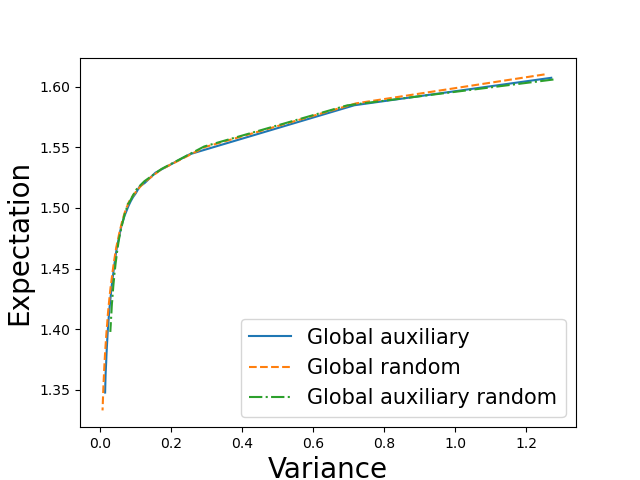}
    \caption*{Dimension 4}
    \end{minipage}
    \begin{minipage}[c]{.49\linewidth}  
    \includegraphics[width=0.99\linewidth]{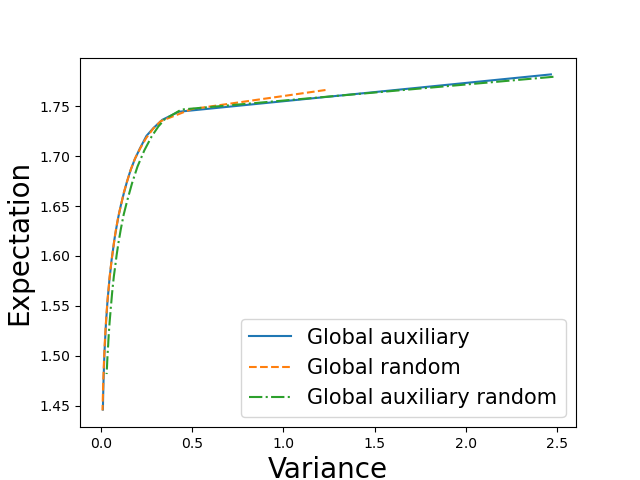}
    \caption*{Dimension 20}
    \end{minipage}     
    \caption{Efficient frontier with dynamic optimization for global random  approach : methods with  stochastic risk coefficients compared to reference calculated with global auxiliary method. }
    \label{fig:dynWeight01Rand}
\end{figure}
As in the analytical case, the randomized version of the global approach has difficulties to represent the whole curve especially in high dimension  as shown on figure \ref{fig:dynWeight01Rand}: in dimension 20, the direct global approach with randomization only gives a part of the curve (permitting to recover the curve only for $\beta$ leading a low variance portfolio) while the auxiliary global random version gives another part of the optimal curve (permitting to recover the curve only for $\beta$ leading a high variance portfolio).
\begin{Remark}
As we sample the parameters to  represent the same part of the curve, the use of a uniform law in the random method to explore the $\beta$ or $\gamma$ values may give too much importance on a given  part of the curve (it depends if a direct or auxiliary approach is used). However results are not improved trying to change the law used to sample the $\beta$ or $\gamma$.
\end{Remark}

Nevertheless, the different methods give very similar results for different values of $\beta/\gamma$  not corresponding to  extreme points of the curve as shown on tables \ref{tab:someBetaDim4DirectConst01}, \ref{tab:someBetaDim4ZhouConst01}, \ref{tab:someBetaDim20DirectConst01}, \ref{tab:someBetaDim20ZhouConst01}.
\begin{table}[H]
    \centering
    \begin{tabular}{|c|c|c|c|c|c|c|} \hline
      $\beta$ &   \multicolumn{2}{c|}{Point by point} &  \multicolumn{2}{c|}{Global} &  \multicolumn{2}{c|}{Global random} \\ \hline
 &  Mean & Variance  & Mean & Variance  & Mean & Variance   \\ \hline
 0.062 & 1.585 & 0.703 & 1.587 & 0.747 & 1.59 & 0.766  \\ \hline
0.821 & 1.503 & 0.08 & 1.497 & 0.073 & 1.497 & 0.072  \\ \hline
5.04 & 1.379 & 0.015 & 1.379 & 0.015 & 1.381 & 0.015  \\ \hline
    \end{tabular}
    \caption{Resolution of \eqref{eq:directNN} for some $\beta$ values with constraint on sum of weights in dimension 4. }
    \label{tab:someBetaDim4DirectConst01}
\end{table}

\begin{table}[H]
    \centering
    \begin{tabular}{|c|c|c|c|c|c|c|} \hline
      $\gamma$ &   \multicolumn{2}{c|}{Point by point} &  \multicolumn{2}{c|}{Global} &  \multicolumn{2}{c|}{Global random} \\ \hline
 &    Mean & Variance  & Mean & Variance  & Mean & Variance   \\ \hline
 9.677 & 1.583 & 0.685 & 1.585 & 0.694 &   1.584 & 0.675  \\ \hline
2.111 & 1.503 & 0.08 & 1.499 & 0.073 & 1.498 & 0.073  \\ \hline
1.478 & 1.38 & 0.015 & 1.386 & 0.019 & 1.378 & 0.015\\ \hline
   \end{tabular}
    \caption{Resolution of \eqref{eq:ZhouForm} for some $\gamma$ values with constraint on sum of weights  in dimension 4.}
    \label{tab:someBetaDim4ZhouConst01}
\end{table}

\begin{table}[H]
    \centering
    \begin{tabular}{|c|c|c|c|c|c|c|} \hline
      $\beta$ &    \multicolumn{2}{c|}{Point by point} &  \multicolumn{2}{c|}{Global} &  \multicolumn{2}{c|}{Global random} \\ \hline
 &  Mean & Variance  & Mean & Variance  & Mean & Variance   \\ \hline
 0.062 & 1.745 & 0.431 & 1.739 & 0.371 & 1.744 & 0.442  \\ \hline
0.821 & 1.65 & 0.126 & 1.651 & 0.11 & 1.651 & 0.11  \\ \hline
5.04 & 1.488 & 0.016 & 1.489 & 0.015 & 1.49 & 0.015  \\ \hline
    \end{tabular}
    \caption{Resolution of \eqref{eq:directNN} for some $\beta$ values with constraint on sum of weights in dimension 20. }
    \label{tab:someBetaDim20DirectConst01}
\end{table}

\begin{table}[H]
    \centering
    \begin{tabular}{|c|c|c|c|c|c|c|} \hline
      $\gamma$ &      \multicolumn{2}{c|}{Point by point} &  \multicolumn{2}{c|}{Global} &  \multicolumn{2}{c|}{Global random} \\ \hline
 &   Mean & Variance  & Mean & Variance  & Mean & Variance   \\ \hline
 9.837 & 1.745 & 0.435 & 1.745 & 0.431 & 1.745 & 0.431  \\ \hline
2.259 & 1.65 & 0.108 &1.651 & 0.11 & 1.653 & 0.11   \\ \hline
1.588 & 1.488 & 0.015 & 1.503 & 0.022 & 1.496 & 0.018  \\ \hline
   \end{tabular}
    \caption{Resolution of \eqref{eq:ZhouForm} for some $\gamma$ values with constraint on sum of weights  in dimension 20. }
    \label{tab:someBetaDim20ZhouConst01}
\end{table}
\subsection{Adding constraints  on the weights}
Investors often impose other constraints on portfolio:
\begin{itemize}
    \item Some are local constraints: the variations of the weights are limited from one step to another in order to face liquidity constraints and to reduce transaction cost:
    \begin{align}
        |\phi_j(t_{i+1}, X_{t_{i+1}})-\phi_j(t_i, X_{t_i})| \le \eta_j,  \mbox{ for } j=1, \dots, d ,\mbox{ and }  i= 0, \dots, N-1 \label{eq:liqConsLoc}
    \end{align}
    \item Some are global constraints : weights are only allowed to stay in a convex compact:
    \begin{align}
    \underline{\phi}_j \le \phi_j(t_i, X_{t_i}) \le  \bar{\phi}_j,  \quad \mbox{ for } j=1, \dots, d.
    \label{eq:liqConsGlob}
    \end{align}
\end{itemize}
We first compare different formulations to solve the problem with the previous constraints.
We test them on  the point by point estimation of the frontier in dimension 4 in the next subsection. 
Then we use the best model to achieve an exhaustive comparison of the point by point and global approaches.
\subsubsection{Presentation of the different models on the point by point approach in dimension 4}
\begin{itemize}
    \item The first model consists in taking the same representation for the weights  as in equation \eqref{eq:NNWeight1}. Then constraints  \eqref{eq:liqConsLoc}, \eqref{eq:liqConsGlob} are imposed by penalization of the objective function and the parameters $\theta$ minimize:
    \begin{align}
    J_1( \hat \phi^\theta) + \beta J_2(\hat \phi^\theta) + \frac{1}{\epsilon} \sum_{j=1}^d \sum_{i=0}^{N-2}  \E \left[(| \hat \phi^\theta_j(t_{i+1}, X^{\hat \phi^\theta}_{t_{i+1}})- \hat \phi^\theta_j(t_i, X^{\hat \phi^\theta}_{t_i})| - \eta_j)^+ \right] +\nonumber\\
    \frac{1}{\epsilon}  \sum_{j=1}^d \sum_{i=0}^{N-1} \E\left( ( \hat \phi^\theta_j(t_i, X^{\hat \phi^\theta}_{t_i}) - \bar{\phi}_j)^{+} + (\underline{\phi}_j - \hat \phi^\theta_j(t_i, X^{\hat \phi^\theta}_{t_i}))^{+} \right)
    \label{eq:optPoids1}
    \end{align}
    where $\epsilon$ is a small penalization parameter.
    \item The second model consists in introducing the variation of weights between two dates similarly as in \cite{fecamp2019risk}. We introduce $\xi^\theta(t,X)$ a neural network taking time and the portfolio value as input  with a tanh  activation function as output so with values in $[-1,1]^d$. The initial portfolio weight is represented by a vector $\tilde \theta$ in $\R^d$ and the weight in the portfolio  at a given date is given by:
    \begin{align}
        \hat \phi^{\theta, \tilde \theta}(t_i, X^{\hat \phi^{\theta, \tilde \theta}}_{t_i}) = \tilde \theta +  \eta \sum_{l=1}^{i} \xi^{\theta}(t_l, X^{\hat \phi^{\theta, \tilde \theta}}_{t_l})
        \label{eq:IncNN}
    \end{align}
    We note $\bar \theta = (\theta, \tilde \theta)$. Local constraints are taken into account in this formulation. It remains to impose global bounds on the weights and the constraint on the summation of the weights. It leads to the minimization in $\bar \theta$ of
    \begin{align}
    J_1( \hat \phi^{ \bar \theta}) + \beta J_2(\hat \phi^{\bar \theta}) + \frac{1}{\epsilon}  \sum_{i=0}^{N-1}  \E \left(| \sum_{j=1}^d \hat \phi^{\bar{\theta}}_j(t_i, X^{\hat \phi^{\bar{\theta}}}_{t_i})- 1|\right) \nonumber\\
    \frac{1}{\epsilon}  \sum_{j=1}^d \sum_{i=0}^{N-1} \E \left( ( \hat \phi^{\bar{\theta}}_j(t_i, X^{\hat \phi^{\bar{\theta}}}_{t_i})- \bar{\phi}_j)^{+} + (\underline{\phi}_j - \hat \phi^{\bar{\theta}}_j(t_i, X^{\hat \phi^{\bar{\theta}}}_{t_i}))^{+} \right)
    \label{eq:optPoids2}
    \end{align}
    \item The third model consist in imposing directly the global constraints at the output of the network \eqref{eq:IncNN}  by introducing:
    \begin{align}
        \hat \phi^{\theta, \tilde \theta}(t_i, X^{\hat \phi^{\theta, \tilde \theta}}_{t_i}) =  ((\tilde \theta +  \eta \sum_{l=1}^{i} \xi^{\theta}(t_l, X^{\hat \phi^{\theta, \tilde \theta}}_{t_l})) \wedge \bar \phi) \vee \underline \phi
        \label{eqIncNNBis}
    \end{align}
    Local and global constraint on the weights are taken into account and it remains to impose that the sum of the weights is equal to one giving the following expression to minimize in $\theta$:
    \begin{align}
    J_1( \hat \phi^{ \bar \theta}) + \beta J_2(\hat \phi^{\bar \theta}) + \frac{1}{\epsilon}  \sum_{i=0}^{N-1}  \E \left(| \sum_{j=1}^d \hat \phi^{\bar{\theta}}_j(t_i, X^{\hat \phi^{\bar{\theta}}}_{t_i})- 1|  \right)
    \label{eq:optPoids3}
    \end{align}  
    \item The fourth and last model consists in using a feedforward network giving as output
    $\zeta^\theta(t,X)$ in $[0,1]^d$  by using a sigmoid activation function at the output of the network. A rescaling is achieved to get the  weights in $[\underline \phi, \bar \phi]$ by
    \begin{align}
   \hat \phi^\theta(t,X)= \underline \phi  +\zeta^\theta(t,X) (\bar \phi - \underline \phi).
   \label{eq:poidsConst}
    \end{align}
    It remains to get an output in the following  hyperplane
     \begin{align}
    1 = \sum_{i=1}^d \hat \phi^\theta(t,X),
    \label{eq:hyperplan}
    \end{align}   
    which can be achieved with the following projection algorithm applied on the network:
    \begin{algorithm}[H]
      \begin{algorithmic}[1]
        \STATE Input : $  \hat \phi^\theta(t,X)$  with values in $[\underline \phi, \bar \phi]$
       \FOR{$i = 1, d$} 
       \STATE $\hat \phi^\theta_i(t,X) = [(\hat \phi^\theta_i(t,X)+(1- \sum_{j=1}^d \hat \phi^\theta_j(t,X))) \wedge \bar \phi_i] \vee \underline \phi_i$
       \ENDFOR
       \STATE Return : $\hat \phi^\theta(t,X)$  satisfying \eqref{eq:hyperplan}
      \end{algorithmic}
      \caption{Projection algorithm applied on the output of the network} \label{algo:projectD}
  \end{algorithm}
   The projections are carried out successively in the different directions. In order to avoid having a preferential direction, it is also possible to randomize the loop of the previous algorithm by performing a random permutation of the different visited dimensions.\\
   The objective function to minimize in  $\theta$ is 
   \begin{align}
    J_1( \hat \phi^\theta) + \beta J_2(\hat \phi^\theta) + \frac{1}{\epsilon} \sum_{j=1}^d \sum_{i=0}^{N-1}  \E \left[(| \hat \phi^\theta_j(t_{i+1}, X^{\hat \phi^\theta}_{t_{i+1}})- \hat \phi^\theta_j(t_i, X^{\hat \phi^\theta}_{t_i})| - \eta_j)^+  \right]
    \label{eq:optPoids4}
    \end{align}
\end{itemize}
\begin{Remark}
In the previous formulations, we have supposed that the initial weights in the portfolio had to be optimized. It is often a data in the problem  and then only weights after the initial date have to be optimized.
\end{Remark}
We test the four previous model on the four dimensional test case described at the beginning of the section.
We impose the additional constraints : the weights cannot vary more than $0.05$ in absolute value  between two rebalancing date and we impose that weights are between $0.1$ and $0.6$. Besides in this case, we impose that initial weights are the same for each asset.\\
We take the following resolution parameters: we take $\epsilon=10^{-4}$, the initial learning rate is set to $10^{-3}$ and the learning rate  decreases to $10^{-5}$ with the gradient iterations. The batch size is equal to $300$ and the number of gradient  iterations equal to $10000$.
For each case, we carry out 4 optimizations and keep the best result (the one with the minimal   objective function).
In the table \ref{tab:consComp}, we give $\E(X_T) - \beta \E((X_T -\E(X_T))^2)$  for the four models for different values of $\beta$.
\begin{table}[H]
    \centering
    \begin{tabular}{|c|c|c|c|c|c|c|c|c|c|} \hline
    $\beta$     &   0  &  0.23 &   0.479  & 0.719 & 0.959  & 1.198 &  1.427  &  1.678 &  2.158\\
    Opt 1     &  1.468 &  1.419 &  1.402 & 1.392 &  1.382&  1.370 & 1.365 & 1.354 &  1.337 \\
    Opt 2     &  1.336 &  1.309 &   1.322 &  1.334 & 1.322&  1.313 &  1.298 &  1.303 & 1.292 \\
    Opt 3     &  1.397 & 1.290 &  1.224 & 1.206 & 1.213 &    1.306 & 1.167 & 1.245&  0.791\\ 
    Opt 4     &  1.474 &  1.426 & 1.409 & 1.398 & 1.387 & 1.378 &1.368 & 1.360&  1.344  \\ \hline
    \end{tabular}
    \caption{Comparison of the  4 models taking into account the constraints  giving  $E(X_T) - \beta E((X_T -E(X_T))^2)$ obtained.}
    \label{tab:consComp}
\end{table}
Models 1 and 4 give the best results. Besides, models 2 and 3 have difficulties to satisfy the constraints: in the following table \ref{tab:consComp}  we give the opposite of the objective function calculated. The difference between results in table below with results in table \ref{tab:consComp} indicate a violation of the constraints.
\begin{table}[H]
    \centering
    \begin{tabular}{|c|c|c|c|c|c|c|c|c|c|} \hline
    $\beta$     &   0  &  0.23 &   0.479  & 0.719 & 0.959  & 1.198 &  1.427  &  1.678 &   2.158   \\
    Opt 1     &  1.427 & 1.371 &1.396 & 1.380 & 1.373 & 1.370 & 1.362 & 1.354&1.337\\
    Opt 2     &  1.335  & 1.307 & 1.320 & 1.331 & 1.318 & 1.313 &1.280 & 1.281 & 1.246 \\
    Opt 3     &  -620 &  -660 &  -942 &-568 & -479 & -1336 &-670&  -611 &  -750 \\
    Opt 4     &  1.474 &1.425 &1.409 & 1.397& 1.386 & 1.378 & 1.368 & 1.360 &1.341\\ \hline
   \end{tabular}
    \caption{Opposite of the objective function.}
    \label{tab:consCompO}
\end{table}
Model 3 seems be to  totally unable to take into account the constraints, while model 2 satisfies globally the constraints.
We may think that the results depends on some hyper parameters, and we test the different algorithms  with a smaller $\epsilon=10^{-5}$. Results are given in table \ref{tab:consCompEps} and \ref{tab:consCompOEps}: they confirm our first results.
\begin{table}[H]
    \centering
    \begin{tabular}{|c|c|c|c|c|c|c|c|c|c|} \hline
    $\beta$     &   0  &  0.23 &   0.479  & 0.719 & 0.959  & 1.198 &  1.427  &  1.678 &  2.158\\
    Opt 1   & 1.465 & 1.415 & 1.395& 1.389 & 1.380 & 1.372 & 1.364 & 1.349 & 1.331 \\ \hline
 Opt 2   & 1.362 & 1.331 & 1.337 & 1.342 & 1.322 & 1.329 & 1.290 & 1.282 &  1.289 \\ \hline
 Opt 3   & 1.431 & 1.370 & 1.308 & 1.259 & 1.192 & 1.189 & 1.163 & 1.245 &  1.198 \\ \hline
 Opt 4   & 1.474 & 1.428 & 1.409 & 1.398 & 1.388 & 1.378 & 1.368 & 1.360 &  1.345 \\ \hline
    \end{tabular}
    \caption{Comparison of the  4 models taking into account the constraints  giving  $E(X_T) - \beta E((X_T -E(X_T))^2)$ obtained using  $\epsilon=10^{-5}$.}
    \label{tab:consCompEps}
\end{table}

\begin{table}[H]
    \centering
    \begin{tabular}{|c|c|c|c|c|c|c|c|c|c|} \hline
    $\beta$     &   0  &  0.23 &   0.479  & 0.719 & 0.959  & 1.198 &  1.427  &  1.678 &   2.158   \\
 Opt 1   & 1.448& 1.414 & 1.394 & 1.377 & 1.380 & 1.363 & 1.353& 1.342 & 1.331 \\ \hline
 Opt 2   & 1.310 & 1.328 & 1.306 & 1.295 & 1.316 & 1.300 & 1.287 & 1.236 & 1.268 \\ \hline
 Opt 3   & -15502 & -12847 & -7690 & -6891 & -14510 & -6284 & -9096 & -8022 & -7221 \\ \hline
 Opt 4   & 1.472 & 1.426 & 1.408 & 1.397 & 1.387 & 1.377 & 1.368 & 1.360 &  1.344 \\ \hline
   \end{tabular}
    \caption{Opposite of the objective function using  $\epsilon=10^{-5}$.}
    \label{tab:consCompOEps}
\end{table}
At last we test the influence of some other parameters: we multiply the learning rate by two, take a batch size of 500,  and a use 20000 gradient iterations. Results  are presented  in tables \ref{tab:consCompModParam}  and \ref{tab:consCompModParam}. Other tests are  not presented here  but, using some smaller learning rates, they  give the same results.
\begin{table}[H]
    \centering
    \begin{tabular}{|c|c|c|c|c|c|c|c|c|c|} \hline
    $\beta$     &   0  &  0.23 &   0.479  & 0.719 & 0.959  & 1.198 &  1.427  &  1.678 &  2.158\\
    Opt 1   & 1.457 & 1.417 & 1.402 & 1.395 & 1.385 & 1.375 & 1.366 & 1.354 &  1.337 \\ \hline
 Opt 2   & 1.348 & 1.374 & 1.348& 1.328& 1.324 & 1.297 & 1.303 & 1.310 & 1.302 \\ \hline
 Opt 3   & 1.2283& 1.369 & 1.346 & 1.224 & 1.198 & 1.195 & 0.989 & 1.204 & 1.083 \\ \hline
 Opt 4   & 1.475 & 1.427& 1.410 & 1.398& 1.388 & 1.378 & 1.368 & 1.360 &  1.345 \\ \hline
    \end{tabular}
    \caption{Comparison of the  4 models taking into account the constraints  giving  $E(X_T) - \beta E((X_T -E(X_T))^2)$ obtained  multiplying learning rates by 2, using a batch size of 500 and 20000 gradient iterations.}
    \label{tab:consCompModParam}
\end{table}
\begin{table}[H]
    \centering
    \begin{tabular}{|c|c|c|c|c|c|c|c|c|c|} \hline
    $\beta$     &   0  &  0.23 &   0.479  & 0.719 & 0.959  & 1.198 &  1.427  &  1.678 &   2.158   \\
    Opt 1   & 1.457 & 1.369 & 1.400 & 1.395& 1.385 & 1.344 & 1.323 & 1.344 &  1.335 \\ \hline
 Opt 2   & 1.297 & 1.280 & 1.346 & 1.304 & 1.265 & 1.296 & 1.270 & 1.237 & 1.230 \\ \hline
 Opt 3   & -284 & -208 & -335 & -309 & -454 & -336 & -185 & -239 &  -148 \\ \hline
 Opt 4   & 1.474 & 1.425 & 1.409 & 1.397 & 1.386 & 1.377 & 1.368 & 1.357 &  1.345 \\ \hline
   \end{tabular}
    \caption{Opposite of the objective function obtained  multiplying learning rates by 2, using a batch size of 500 and 20000 gradient iterations.}
    \label{tab:consCompOModParam}
\end{table}
Once again models 1 and 4 are the best to respect the constraints and  results for model 3 (always very bad) vary a lot with the hyper parameters taken.
\subsubsection{Point by point approximation for efficient frontier}
\label{sec:ptbyptCons}
We keep model 1 and model 4 to test them further in different dimensions.
We suppose that the weights are between $0.5 p_{Init}$ and $2 p_{Init}$ where $p_{Init}=\frac{1}{d}$ is the initial equal weight used in the portfolio.
Local constraints are kept as in the previous subsection with variations below $0.05$ in absolute value between two rebalancing date.
Optimization parameters are:
\begin{itemize}
    \item Model 1: initial learning rate equal to $\frac{0.8 10^{-3}}{d}$ and linearly decreasing to $\frac{0.8 10^{-4}}{d}$ with gradient iterations,
    \item Model 4: initial learning rate equal to $10^{-3}$ and linearly decreasing to $10^{-5}$.
\end{itemize}
\begin{Remark}
Model 1 has a convergence more sensitive than Model 4 to the learning rate depending on the dimension. We found that such a scaling gives good results.
\end{Remark}
25000 iterations of gradient are used. The batch size is equal to $100$ to reduce the computational time as we found that this is sufficient. In dimension 4, the efficient frontier  is estimated using $40$ discretization points. In dimension 20, 40 discretization points are used with model 1 while 16 points are used only with model 4: it is important to notice that the projection algorithm leads to a computational time  far more important as the dimension of the problem increases.
\begin{figure}[H]
    \centering
    \begin{minipage}[c]{.49\linewidth}
    \includegraphics[width=0.99\linewidth]{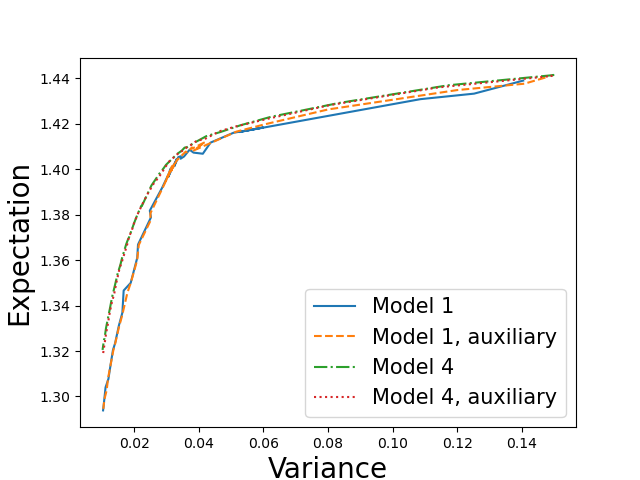}
    \caption*{Dimension 4}
    \end{minipage}
    \begin{minipage}[c]{.49\linewidth}  
    \includegraphics[width=0.99\linewidth]{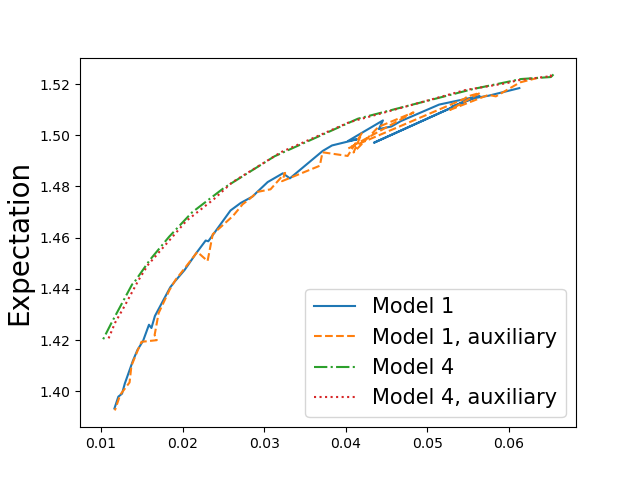}
    \caption*{Dimension 20}
    \end{minipage}     
    \caption{Comparing model 1 and 4 taking into account constraints in a point by point approximation of the efficient frontier.}
    \label{fig:MarkovitzPoids01ConstraintsCompPenalNew}
\end{figure}
In figure \ref{fig:MarkovitzPoids01ConstraintsCompPenalNew}, we compare model 1 and 4 to calculate the efficient frontier using the direct or auxiliary problem. Direct and auxiliary problems give the same frontier for both methods and model 4 is clearly superior to model 1.
In the following model 4 will be used to deal with the global/local constraints. 

\subsubsection{Global optimization and comparison with point by point optimization of the efficient frontier.}
We keep the constraints defined in  section \ref{sec:ptbyptCons}.
The model 4 can be extended easily to estimate the frontier globally: the network $\zeta^\theta$  with parameters $\theta$ is now  a function of $t, X, \beta$  still with values in $[0,1]^d$.
The weight approximation is carried out by:
\begin{align}
  \hat \phi^{\theta;\beta}(t,X, \beta)= \underline \phi  +\zeta^\theta(t,X, \beta) (\bar \phi - \underline \phi).
  \label{eq:globalConst}
\end{align}
The projection algorithm \eqref{eq:hyperplan} is used and the objective function \eqref{eq:solvMeanVarBeta} is replaced by:
\begin{align}
   \theta^{*} = & \argmin_{\theta} \E[ -\E[X_T^{\hat \phi^{\theta; \hat \beta}}|\hat \beta] + \hat \beta \E[ (X_T^{\hat \phi^{\theta; \hat \beta}}+ \E[X_T^{\hat \phi^{\theta; \hat \beta}}| \hat \beta])^2|\hat \beta]] + \nonumber \\
   & \frac{1}{\epsilon} \sum_{j=1}^d \sum_{i=0}^{N-2}  \E[| \hat \phi^\theta_j(t_{i+1}, X^{\hat \phi^\theta; \hat \beta}_{t_{i+1}}, \hat \beta )- \hat \phi^\theta_j(t_i, X^{\hat \phi^\theta; \hat \beta}_{t_i}, \hat \beta )| - \eta_j)^+ ]
   \label{eq:optConstGlob}
\end{align}
Of course a similar objective function can be written for the auxiliary problem.\\
We only present the results obtained using the deterministic (for $\beta$ and $\gamma$)  global method as the randomized  approach  tends to represent only a part of the curve or is suboptimal as seen on the case without constraints.
As before, in dimension 4, $40$ points are used to estimate the efficient frontier while only $16$ are used in dimension $20$.
\begin{figure}[H]
    \centering
    \begin{minipage}[c]{.49\linewidth}
    \includegraphics[width=0.99\linewidth]{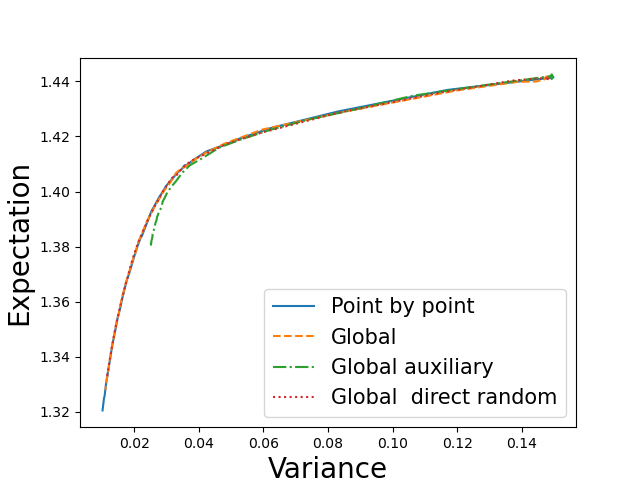}
    \caption*{Dimension 4}
    \end{minipage}
    \begin{minipage}[c]{.49\linewidth}  
    \includegraphics[width=0.99\linewidth]{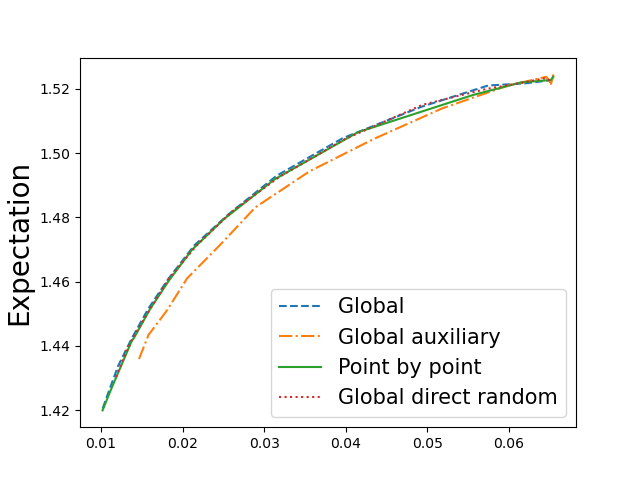}
    \caption*{Dimension 20}
    \end{minipage}     
    \caption{Point by point versus global efficient frontier estimation.}
    \label{fig:FrontierNewSchemeCompPontbPointGlobal}
\end{figure}
On figure \ref{fig:FrontierNewSchemeCompPontbPointGlobal}, we plot the efficient frontier obtained by point by point and global estimation.
The point by point calculation and the direct global calculations (deterministic and randomized version) give the same curves in dimension 4 and 20.
The global auxiliary approach gives a suboptimal curve in dimension 4 and 20 for its deterministic version while the randomize version gives solution with a flat variance in both cases so is not reported.
\begin{Remark}
In  the whole section, we have added local constraints in the strategies. It is also possible to remove the local constraints and take into account the transaction costs directly in the dynamic of the asset. Supposing that half of spread bid ask for asset $i$ is given by $p_i$, and still supposing that we deal with the discrete-time  optimization \eqref{eq:conPorfolioDis2}, then the final value of the assets is given with the convention  $\phi_{-1}=0$ by
\begin{align}
    X_T^{\phi} = X_0 + \sum_{i=0}^{N-1} \phi_i X_{t_i}^{\phi}. \frac{S_{t_{i+1}}- S_{t_i}}{S_{t_i} }
    + \sum_{i=0}^{N-1}  \sum_{j=1}^d  | \frac{\phi_{i,j} X_{t_i}^{\phi}}{S_{t_i,j}} - \frac{\phi_{i-1,j} X_{t_{i-1}}^{\phi}}{S_{t_{i-1},j}} | p_j
\end{align}
Now the control depends not only on the wealth but also on $S_t$ and the weight at the previous date and they have to be included in the state at the input of the network.
\end{Remark}

\section{Time discrete optimization under an Heston model}
We now suppose that the assets follow the Heston model \cite{heston1993closed}:
\begin{align}
\label{eq:Heston}
    dS_t =&  \mu S_t dt+ \sqrt{V_t} S_t dW^1_t, \nonumber\\
    dV_t = & \kappa (\bar V - V_t) dt + \bar \sigma  \sqrt{ V_t} d W^2_t
\end{align}
where $S_t$, $V_t$ with values in $\R^d$, $(W^1_t,W^2_t)$ is a vector of $2d$ Brownian correlated with a correlation matrix $\rho$.
As there is no analytic solution to equation  \eqref{eq:Heston}, we rely on  a Milstein scheme on the volatility \cite{kahl2006fast}  to assure positivity of the volatility under the Feller condition $2 \kappa \bar V  \le  \bar{\sigma}^2$.\\
We are only interested in the discrete problem with no short selling and no borrowing : weights are all in $[0,1]$ and the summation of the weights is equal to one.
This problem is interesting  as the state of the problem now involve not only the wealth $X_t$ but also the variance $V_t$, such that not only the control but also  the global state has a dimension increasing with the number of assets in the portfolio.\\
We first deal with the case without additional constraints and then the case with  local and global constraints.
For all optimizations, we take a learning rate  initially equal to  $10^{-3}$ and linearly decreasing to $10^{-4}$ with stochastic gradient iterations.
We take $25000$ gradient iterations except when specified. The batch size is equal to $100$.
\subsection{No additional constraints}
\begin{figure}[H]
    \centering
    \begin{minipage}[c]{.49\linewidth}
    \includegraphics[width=0.99\linewidth]{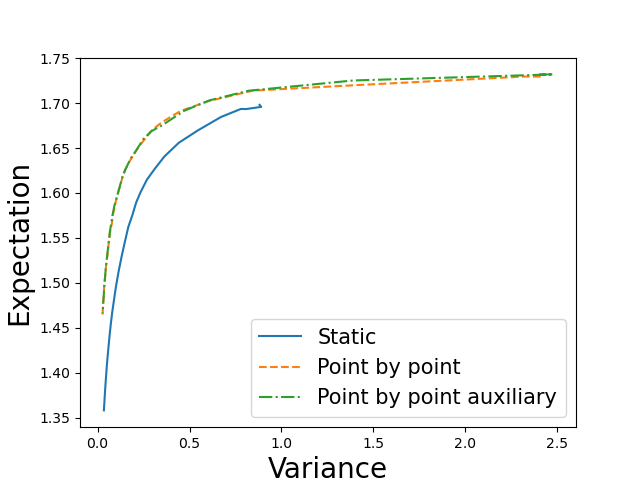}
    \caption*{Point by point}
    \end{minipage}
    \begin{minipage}[c]{.49\linewidth}  
    \includegraphics[width=0.99\linewidth]{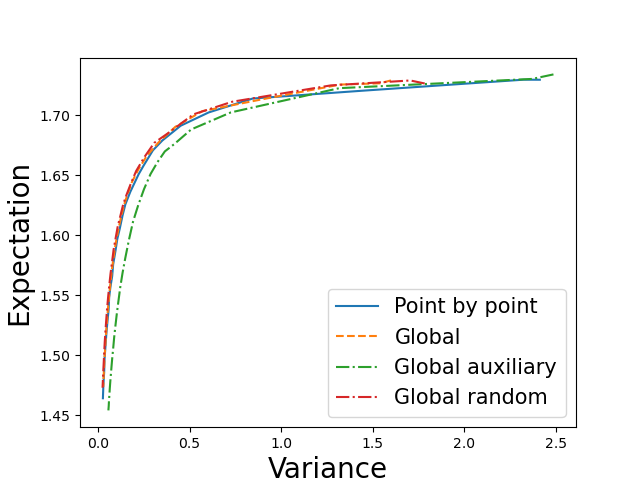}
    \caption*{Point by point versus global}
    \end{minipage}     
    \caption{Results in dimension 10 for the Heston model without additional constraints.}
    \label{fig:Heston10D}
\end{figure}

Using the previous algorithm equation \eqref{eq:NNWeight1} and \eqref{eq:NNWeight1Glob} are now replaced  by 
\begin{align}
  \hat \phi^\theta(t,X, V_1, \dots, V_d) = \frac{\zeta^{\theta}(t,X, V_1, \dots, V_d)}{\sum_{i=1}^d \zeta^{\theta}_i(t,X, V_1, \dots, V_d)}  
  \label{eq:NNWeightheston}
\end{align}
for the point by point approximation and by
\begin{align}
  \hat \phi^{\theta;\beta}(t,X, V_1, \dots, V_d, \beta) = \frac{\zeta^\theta(t,X, V_1, \dots, V_d, \beta)}{\sum_{i=1}^d\zeta^\theta_i(t,X, V_1, \dots, V_d, \beta)}  
  \label{eq:NNWeighthestonBeta}
\end{align}
in the global method where $V_j$ is the variance of the asset $j$.\\
In the whole section we still take $T=10$ years and a rebalancing every month.
In dimension 4 we take the following parameters:
$$\mu = (0.01  ,0.0225, 0.035,  0.0475)^T,$$   $\bar V$  is taken equal to  $V_0$ and given by $(0.0025, 0.01 ,  0.0225 , 0.04)^T$. Volatility of the variance  $\bar \sigma$  is given by the diagonal matrix with coefficients $(0.05, 0.1 , 0.15, 0.2 )$. All the  $\kappa$ values are equal to $0.5$.  The initial asset values are all equal to one.  The correlation matrix associated with $ (W^{1}_{t,1}, \dots, W^{1}_{t,d}, W^{2}_{t,1}, \dots,  W^{2}_{t,d})$ is given by 
{\small
\begin{align*}
    \rho = \left (
    \begin{array}{cccccccc}
1.     &    -0.383& 0.378& -0.324& -0.751 &-0.110 &
   0.272 & 0.465  \\
 -0.383 & 1.    &     -0.938  & -0.411 & -0.053&  0.276 &
  -0.226&   -0.349 \\
  0.378& -0.938 &  1.  &         0.145&   -0.051 & -0.398& 
   0.249 &  0.401 \\
 -0.324 & -0.411 &  0.145&    1.   &        0.655 &  0.329 &
  -0.264& -0.048 \\
 -0.751 & -0.053 & -0.051 & 0.655 &  1. &         -0.172 &
  -0.464 & -0.105 \\
 -0.110 &  0.276& -0.398 & 0.329 & -0.172&   1.& 
   0.044 & -0.348 \\
  0.272 &-0.226 &   0.249 & -0.264& -0.464 &  0.044 &
   1.    &     -0.580 \\
 0.465&   -0.349 &  0.401& -0.048 & -0.105& -0.348& 
  -0.580 &  1.    
 \end{array}
    \right).
\end{align*}
}
In dimension 10 a similar test case is created with quite high correlations.
All curves are plotted using 30 points.
The convergence of the "global auxiliary" being very slow, $50000$ gradient iterations have been used specially for the curves "global auxiliary". 
\begin{figure}[H]
    \centering
    \begin{minipage}[c]{.49\linewidth}
    \includegraphics[width=0.99\linewidth]{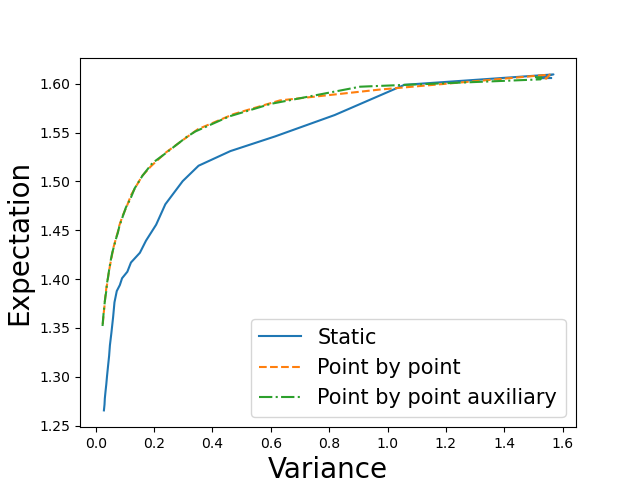}
    \caption*{Point by point}
    \end{minipage}
    \begin{minipage}[c]{.49\linewidth}  
    \includegraphics[width=0.99\linewidth]{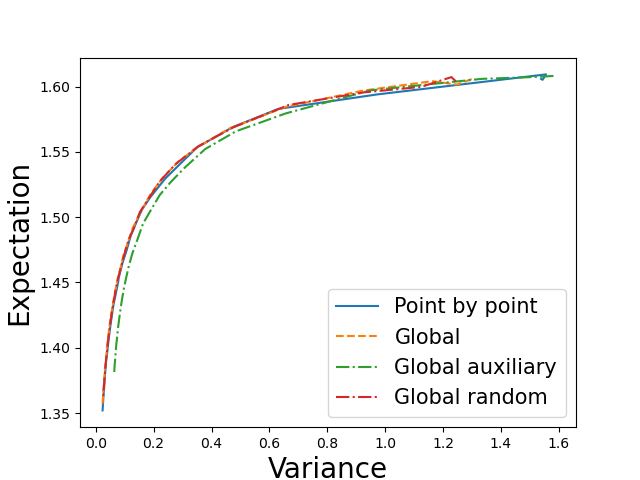}
    \caption*{Point by point versus global}
    \end{minipage}     
    \caption{Results in dimension 4 for the Heston model without additional constraints.}
    \label{fig:Heston4D}
\end{figure}
In figure \ref{fig:Heston4D}, we first plot in dimension 4 the efficient frontier obtained by direct optimization and using auxiliary equations. Both approaches give the same curve, well above the static curve which surprisingly doesn't seem to be very well estimated (by a point by point approach) as convexity of the curve is not totally respected.
In figure \ref{fig:Heston4D}, we also compare global estimations to point by point estimations: using the direct approach  (deterministic and randomized version) we get the same curve as the one given by  the point by point estimation. Using the auxiliary version with the global approach, we get similar results as in the Black-Scholes case:  the deterministic version gives a suboptimal curve while the randomized version is not reported as it gives bad results.\\
In figure \ref{fig:Heston10D}, we give the same results in dimension 10. We get very similar results, except that the static point by point  optimization curve is more realistic.
\subsection{Imposing additional local and global constraints}
We keep the constraints used in the Black-Scholes model in section \ref{sec:ptbyptCons}.
The equation \eqref{eq:poidsConst} is modified taking into account the fact that the state includes the variance of the different assets for the point by point optimization and the equation \eqref{eq:globalConst} is  modified in the same way for global optimization.
\begin{figure}[H]
    \centering
    \begin{minipage}[c]{.49\linewidth}
    \includegraphics[width=0.99\linewidth]{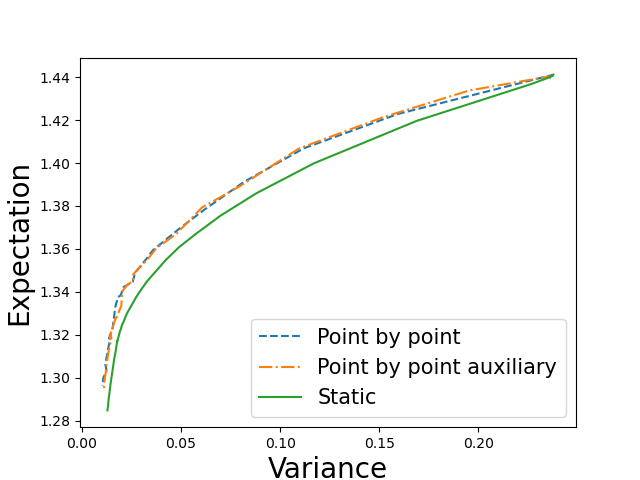}
    \caption*{Point by point}
    \end{minipage}
    \begin{minipage}[c]{.49\linewidth}  
    \includegraphics[width=0.99\linewidth]{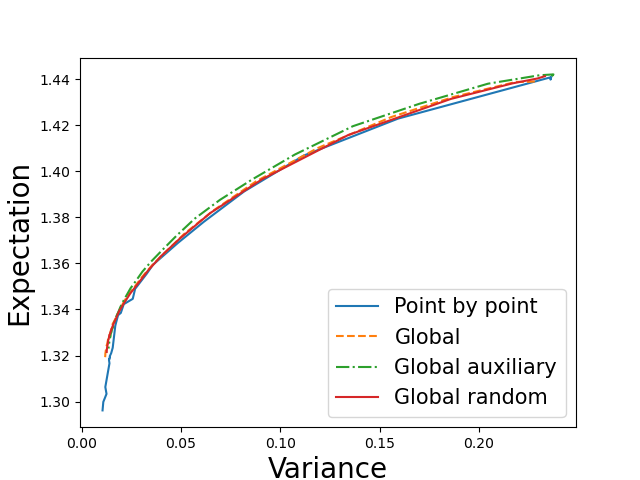}
    \caption*{Point by point versus global}
    \end{minipage}     
    \caption{Results in dimension 4 for the Heston model with additional constraints.}
    \label{fig:Heston4DConst}
\end{figure}
Figure \ref{fig:Heston4DConst} seems to indicate that we are able to recover the frontier with a good accuracy both by the point by point and the global method.
\begin{figure}[H]
    \centering
    \begin{minipage}[c]{.49\linewidth}
    \includegraphics[width=0.99\linewidth]{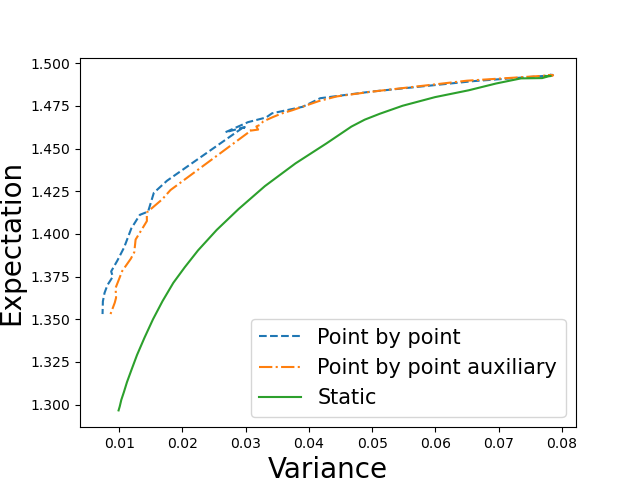}
    \caption*{Point by point}
    \end{minipage}
    \begin{minipage}[c]{.49\linewidth}  
    \includegraphics[width=0.99\linewidth]{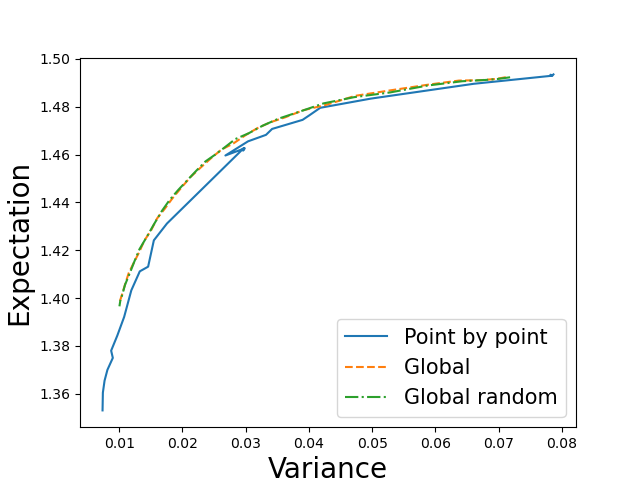}
    \caption*{Point by point versus global}
    \end{minipage}     
    \caption{Results in dimension 8 for the Heston model with additional constraints.}
    \label{fig:Heston8DConst}
\end{figure}
In higher dimension (for example in figure \ref{fig:Heston8DConst} in dimension 8), the convergence is harder to achieve and small oscillations appear for the point by point approach. In this case, the global auxiliary results are not reported as the curve obtained is not satisfactory. Globally, the global version on the direct  approach appears to be the most effective.

\section{Mean-CVaR risk optimization}
In the whole section we suppose that the assets follow a Black-Scholes model given by equation \eqref{eq:BS}.
As the Mean-Variance case penalizes both gains and losses, practitioners prefer to use some downside risk only penalizing the losses (or small gains). In this part we focus on the Mean-CVaR criterion.\\
We first recall the VaR definition: 
if $X$ is a distribution of gain with a continuous cumulative distribution $F_X(x)$ and  $\alpha \in ]0,1[$, we note
\begin{align}
    VaR(X, \alpha) =  \min( y \in \R / P(-X \le  y)  \ge \alpha)
\end{align}
where $P(-X \le x)$ is the probability  that the random variable   $-X$ is below $x$, so that 
$P(-X \le x) =  F_{-X}(x)$.\\
CVaR is then the average loss conditionally to the fact that the losses are above the VaR:
\begin{align}
CVaR(X, \alpha)=   \E[ -X / -X \ge VaR(X, \alpha)]
\end{align}
The VaR criterion is not convex but the CVaR is convex \cite{rockafellar2000optimization}.
The Mean-CVaR problem consists in finding strategies adapted to the available information and minimizing
\begin{align}
    (J_1(\phi), J_2(\phi)) = ( -\E[ X_T^{\phi}], CVaR(X_T^\phi -X_0, \alpha)).
    \label{eq:paretoMCVar}
\end{align}
where  $X_T^{\phi}$ is the value of portfolio managed with strategy $\phi$ and given by \eqref{eq:conPorfolioDis2}.\\
As for the Mean-Variance case, an admissible strategy $\phi^{*}$  is efficient if there is no other admissible strategy $\psi$ such that
\begin{align*}
    J_1(\psi) \le J_1(\phi^{*}), \quad  J_2(\psi) \le J_2(\phi^{*})
\end{align*}
with at least one of the last inequality being strict. The  set of efficient points $( J_1(\phi^{*}), J_2(\phi^{*})) $  defines the efficient frontier.\\
As recalled in the introduction, the continuous optimization problem is not well posed and we will only focus on the  discrete-time optimization problem.\\
It is also possible to recast the problem as minimizing the CVaR under a constraint that the expected value of the gains is above a threshold $M$:
\begin{align}
\min_{\phi \in \mathcal{L}_{\mathcal{F}_t}^2(0,T,\R^d)}  CVar(X_T^{\phi}-X_0) \\
\mbox{ with } E[ X_T^{\phi}] \ge M
\label{eq:refCVaRNC}
\end{align}
so that using \cite{rockafellar2000optimization} representation, 
\begin{align}
\min_{ y, \phi \in \mathcal{L}_{\mathcal{F}_t}^2(0,T,\R^d)}  \E[ y + \frac{1}{1- \alpha}( -X_T^{\phi}+X_0-y)^{+}]\\
\mbox{ with } E[ X_T^{\phi}] \ge M
\label{eq:rockfellarForm}
\end{align}
and when optimality is reached,  $y$ corresponds to the VaR  of the portfolio.\\
Conventional numerical methods generally prefer this formulation as if $y$ is fixed, the problem can be solved by dynamic programming. Then combining a gradient descent method with this optimization where $y$ is fixed, it possible to optimize the portfolio. \\
The interest of this formulation is not obvious with neural network problem as  it adds a fictitious dimension on the problem and it turns out that using this formulation goes not give good results such that we don't report them in the article.\\
In the sequel use the same formulation as in the Mean-Variance case: the efficient
 frontier can be calculated by minimizing \eqref{eq:minimPro} using  the definition
 \eqref{eq:paretoMCVar} for a fixed $\beta$ and let $\beta$ vary to describe the frontier.\\
 As in the Mean-Variance case, it is possible to use a neural network to optimize the strategies by minimizing
 equation \eqref{eq:directNN} for a point by point approximation of the frontier or minimizing the single problem
 \eqref{eq:solvMeanVarBeta} to learn the global curve.\\
 In all the tests, the parameters of the model are the same as in  section \ref{sec:DisMeanVar}. As before, $T=10$ years and rebalancing is achieved every month.\\
 As for the convergence of the stochastic gradient, the learning rate starts at $10^{-4}$ and decreases linearly with gradient iterations to $10^{-5}$. The number of stochastic gradient iterations is fixed at $15000$. The batch size is chosen equal to $2000$ to have  a correct  assessment  of the CVaR.  While calculating the curve point by point or globally with deterministic $\beta$,  the following values $\beta_i= (i2/K)^2$, $i=0,\dots, K-1$ are used. The value for $K$ depends on the case. When  a randomization of the parameter $\beta$ is used,  $\hat \beta \sim \beta_{K-1} U^2$ where $U$ is a uniform random variable on $[0,1]$.
 Other distributions have been tested trying to increase the density of  points where the portfolio is more risky, but the global curve was not as good as with the uniform distribution. 
 
 \subsection{Not short selling, no borrowing}
\begin{figure}[H]
    \begin{minipage}[c]{.49\linewidth}  
    \includegraphics[width=0.99\linewidth]{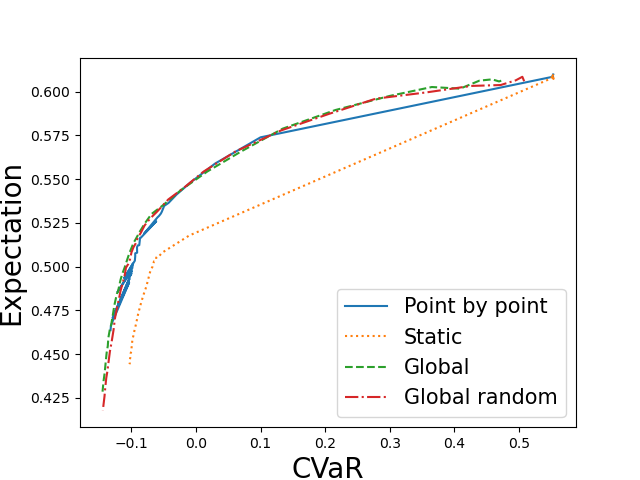}
    \caption*{$\alpha=0.9$}
    \end{minipage}
    \begin{minipage}[c]{.49\linewidth}  
    \includegraphics[width=0.99\linewidth]{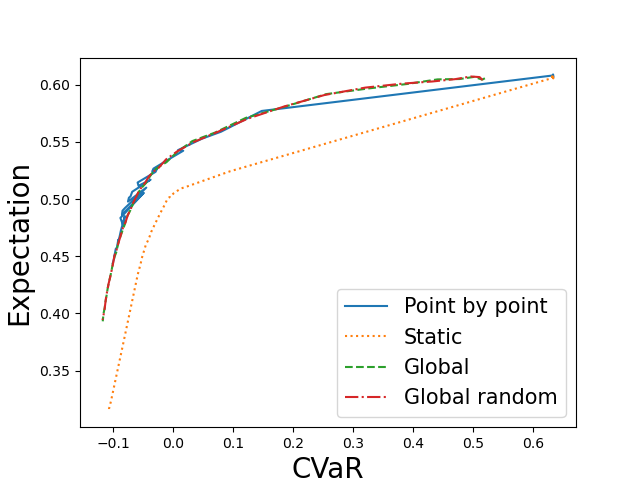}
    \caption*{$\alpha=0.95$}
    \end{minipage}
    \caption{Mean-CVaR optimization  comparison between point by point optimization with global optimization with deterministic and stochastic $\beta$  in dimension 4.}
    \label{fig:CVarComLocalGlobFit4D}
\end{figure}

\begin{figure}[H]
    \begin{minipage}[c]{.49\linewidth}  
    \includegraphics[width=0.99\linewidth]{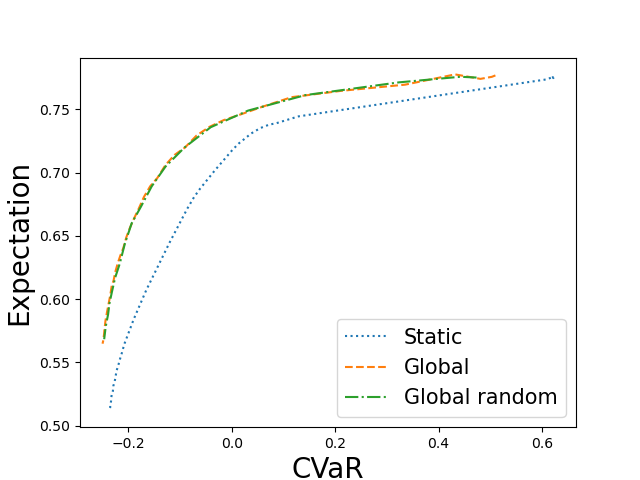}
    \caption*{$\alpha=0.9$}
    \end{minipage}
    \begin{minipage}[c]{.49\linewidth}  
    \includegraphics[width=0.99\linewidth]{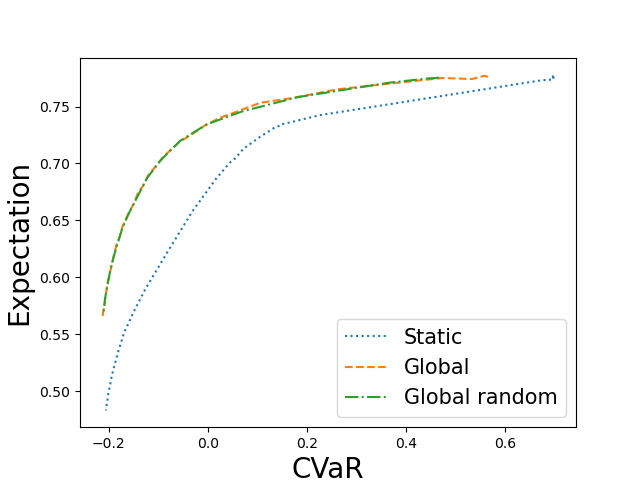}
    \caption*{$\alpha=0.95$}
    \end{minipage}
    \caption{Mean-CVaR optimization  comparison between static  optimization and global optimization with deterministic and stochastic $\beta$  in dimension 20.}
    \label{fig:CVarComLocalGlobFit20D}
\end{figure}

 Similarly to the Mean-Variance we can impose constraints on the portfolio: using the weights formulation \eqref{eq:NNWeight1} while optimizing \eqref{eq:minimPro}, the weights formulation \eqref{eq:NNWeight1Glob} optimizing \eqref{eq:solvMeanVarBeta}, it is possible to impose that all weights are positive, between 0 and 1 and with sum equal to one.\\
 We train the network and plot the resulting efficient frontier using $40$ points first in dimension 4 on figure \ref{fig:CVarComLocalGlobFit4D}. The global approaches  with deterministic and stochastic $\beta$ give the same curve.  The point by point curve appears to be oscillating and may not be very accurate.
In dimension 20, results with point by point approximation are not reported as  they are not usable. Only global solution with deterministic  and stochastic $\beta$ are given on figure \ref{fig:CVarComLocalGlobFit20D}.
For the Mean-CVaR criterion, it seems that the global approximation gives the best results.
 \subsection{Adding other constraints}
\begin{figure}[H]
    \begin{minipage}[c]{.49\linewidth}  
    \includegraphics[width=0.99\linewidth]{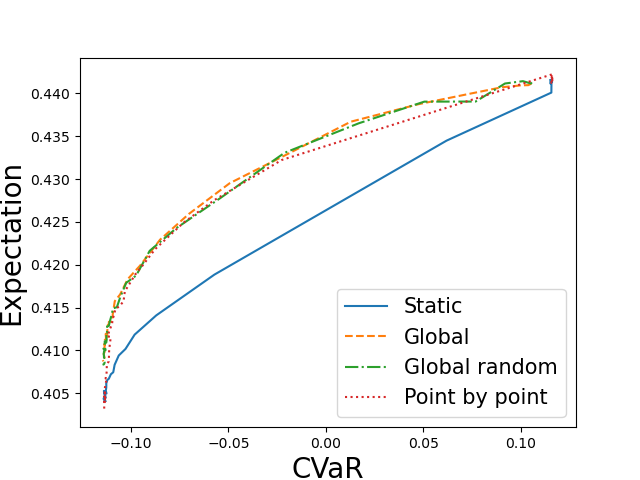}
    \caption*{$\alpha=0.9$}
    \end{minipage}
    \begin{minipage}[c]{.49\linewidth}  
    \includegraphics[width=0.99\linewidth]{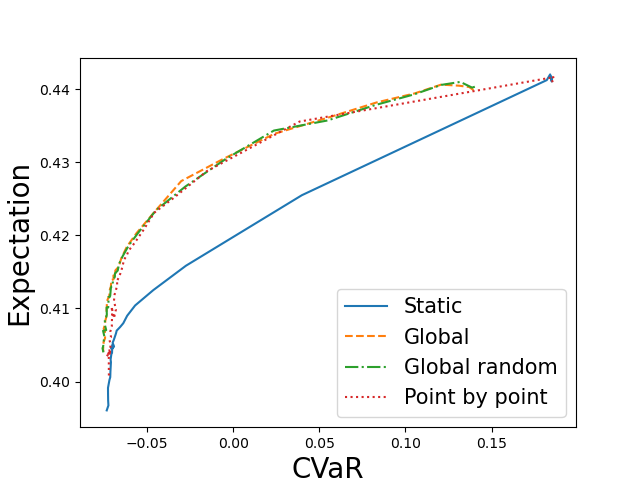}
    \caption*{$\alpha=0.95$}
    \end{minipage}
    \caption{Mean-CVaR in dimension 4 with additional constraints on weights : global, point by point, and static estimations.}
    \label{fig:CVarGlobFit4DConstraints}
\end{figure}

\begin{figure}[H]
    \begin{minipage}[c]{.49\linewidth}  
    \includegraphics[width=0.99\linewidth]{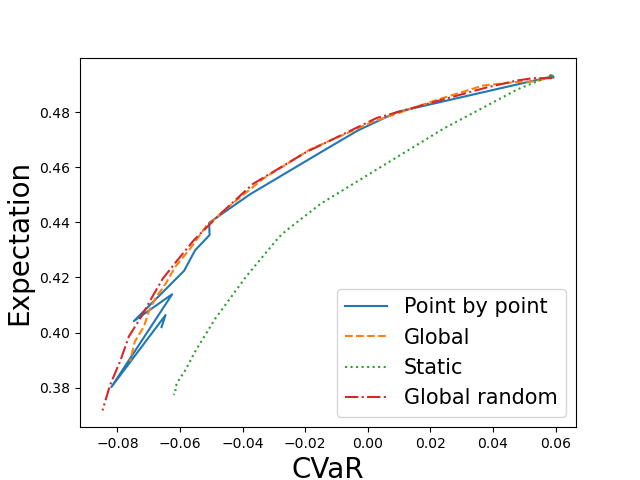}
    \caption*{$\alpha=0.9$}
    \end{minipage}
    \begin{minipage}[c]{.49\linewidth}  
    \includegraphics[width=0.99\linewidth]{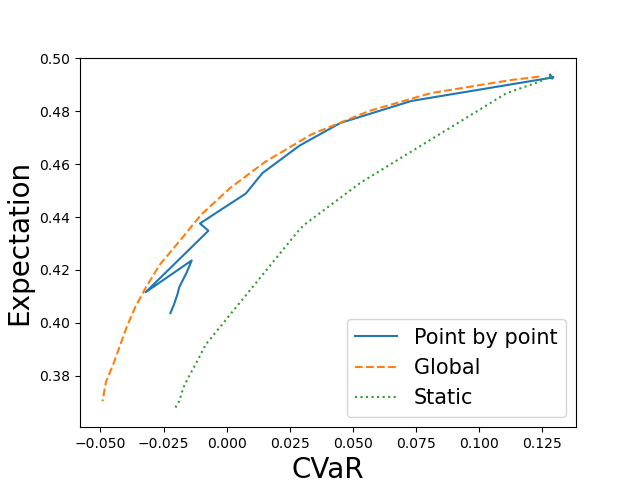}
    \caption*{$\alpha=0.95$}
    \end{minipage}
    \caption{Mean CVaR in dimension 8 with additional constraints on weights : global, point by point, and static estimations.}
    \label{fig:CVarGlobFit8DConstraints}
\end{figure}

\begin{figure}[H]
\centering
\includegraphics[width=0.49\linewidth]{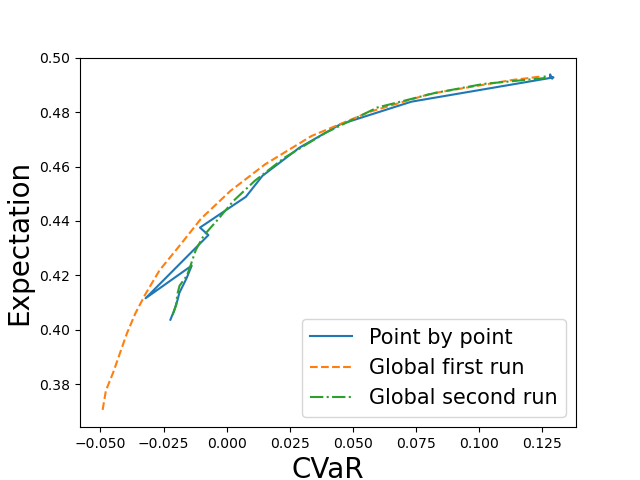}
\caption{ \label{fig:CVarGlobFit8DConstraints2Run} Different Mean CVaR runs with deterministic global method compared to point by point estimation with $\alpha=0.95$.}
\end{figure}

Using equation \eqref{eq:poidsConst} with algorithm \ref{algo:projectD} in equation \eqref{eq:optPoids4}, or equation \eqref{eq:globalConst} for the weights while optimizing \eqref{eq:optConstGlob}, it is possible to add the constraints given by equations \eqref{eq:liqConsLoc} and \eqref{eq:liqConsGlob}. The parameters for the  additional constraints are the same as in section \ref{sec:ptbyptCons}. Results are reported in dimension 4 and only in dimension 8 as computing cost grows significantly with the dimension of the problem. Curves are reconstructed with $20$ points. On figure \ref{fig:CVarGlobFit4DConstraints}, we plot the curves obtained with a static optimization, and the dynamic optimization with the global approaches and the point by point approach. The global approaches seem to give slightly better results.
In higher dimension, the point by point approach appears to give oscillations. It turns out that  the best solution is given by the global approach with random $\beta$ coefficients as shown on figure \ref{fig:CVarGlobFit8DConstraints}.
For example, with  $\alpha=0.95$, the global resolution can lead to a suboptimal curve as shown on figure \ref{fig:CVarGlobFit8DConstraints2Run}. The optimizer is trapped in a local minimum away from the solution and point by point estimations oscillates between the two curves. 
So it seems that with or without additional constraints, the global approach  generally gives  more satisfactory curves than the point by point approach and  it is smoother by construction.
\section{Conclusion}
We have shown that neural network can be used to efficiently compute the Markovitz frontier with or without weight constraints.\\
In the Mean-Variance case, all methods  seem to approximate the Markovitz frontier very well for portfolios with not a too high or too low variance.  Global random methods may have difficulties to either catch very low variance portfolios or very high variance portfolios depending on the formulation taken. When there are local and global constraints on the weights, the new projection algorithm presented in the article has to be used to get accurate results. Nevertheless, the  accuracy of the curve obtained depends on the asset model and the dimension of the model. Therefore we cannot say which formulation (direct \eqref{eq:directNN} or auxiliary \eqref{eq:ZhouForm}) is the best and which resolution method "global" or "point by point" has to be used to get the best results.\\
In the Mean-CVaR case, the batch size to estimate the CVaR has to be very high, and the computational time is currently a limiting factor.
The global methods appears to be more effective than point by point methods: imposing no borrowing, and no short selling, the curves obtained seem to be realistic.
Adding additional constraints increasing a lot the computational time, the number of assets has to be limited and the solver can be trapped by local minima rather different depending on the run: it gives high oscillations with point by point methods and sometimes  converges to a sub-optimal curve with global methods.\\
All the results in this article have been obtained using classical feedforward networks: some work on the structure of the networks used could perhaps improve the results in the difficult Mean-CVaR case.\\
Another interesting approach would be to directly optimize the expected gain with  given bounds on the variance of the portfolio during the whole period or at some given dates:
\begin{align*}
    \max_{\xi} \quad &  \E[ X^\xi_T] \\
     & \E[(X^\xi_{T_i} -\E[X^\xi_{T_i}])^2] \le M, \mbox{ for } i=1, \dots, N,
\end{align*}
where $T_1 < \dots  < T_N=T$, leading to a better control of the risk of the portfolio during the whole period. The resolution of this problem using Neural Networks is a current subject of research.
\printbibliography

\end{document}